\documentclass[submission,copyright,creativecommons, noncommercial, sharealike]{eptcs}
\providecommand{\event}{QPL 2023} 

\usepackage{iftex}

\ifpdf
  \usepackage{underscore}         
  \usepackage[T1]{fontenc}        
\else
  \usepackage{breakurl}           
\fi

\usepackage[english]{babel}
\usepackage{amsmath}								
\usepackage{amsthm} 
\usepackage{amssymb}								
\usepackage{textgreek} 
\usepackage{euler} 
\usepackage{mathpartir} 
\usepackage{allrunes} 
\usepackage{physics} 
\usepackage[scr=rsfso, bb=boondox]{mathalpha}
\usepackage{stmaryrd} 
\usepackage{tikzit} 
\usepackage{subcaption} 
\usepackage{graphics}
\usepackage{mathtools} 
\usepackage{comment} 
\usepackage{amsthm}
\usepackage{quantikz}

\usetikzlibrary{decorations.pathmorphing}
\usetikzlibrary{fadings}
\usetikzlibrary{decorations.pathreplacing}
\usetikzlibrary{decorations.markings}

\tikzstyle{box}=[shape=rectangle, text height=1.5ex, text depth=0.25ex, yshift=0.5mm, fill=white, draw=black, minimum height=5mm, yshift=-0.5mm, minimum width=5mm, font={\small}]
\tikzstyle{mbox}=[shape=rectangle, text height=1.5ex, text depth=0.25ex, yshift=0.5mm, fill=white, draw=black, minimum height=1.5cm, yshift=-0.5mm, minimum width=0.75cm, font={\small}]
\tikzstyle{letter}=[shape=rectangle, text height=1.5ex, text depth=0.25ex, yshift=0.5mm, fill=white, draw=none, minimum height=.5mm, yshift=-0.5mm, minimum width=.5mm, font={\small}]
\tikzstyle{Z}=[inner sep=0mm, minimum size=2mm, shape=circle, draw=black, fill={rgb,255: red,216; green,248; blue,216}]
\tikzstyle{Zp}=[minimum size=5mm, font={\footnotesize\boldmath}, shape=rectangle, rounded corners=2mm, inner sep=0.2mm, outer sep=-2mm, scale=0.8, tikzit shape=circle, draw=black, fill={rgb,255: red,216; green,248; blue,216}, tikzit draw=blue]
\tikzstyle{X}=[Z, shape=circle, draw=black, fill={rgb,255: red,232; green,165; blue,165}]
\tikzstyle{Xp}=[Zp, tikzit shape=circle, tikzit draw=blue, fill={rgb,255: red,232; green,165; blue,165}, font={\footnotesize\boldmath}]
\tikzstyle{Y}=[Z, shape=circle, draw=black, fill={rgb,255: red,165; green,200; blue,232}]
\tikzstyle{Yp}=[Zp, tikzit shape=circle, tikzit draw=blue, fill={rgb,255: red,165; green,200; blue,232}, font={\footnotesize\boldmath}]
\tikzstyle{T}=[Z, shape=circle, draw=black, fill={rgb,255: red,198; green,165; blue,232}]
\tikzstyle{Tp}=[Zp, tikzit shape=circle, tikzit draw=blue, fill={rgb,255: red,198; green,165; blue,232}, font={\footnotesize\boldmath}]
\tikzstyle{H}=[fill=yellow, draw=black, shape=rectangle, inner sep=0.6mm, minimum height=1.5mm, minimum width=1.5mm]
\tikzstyle{Zb}=[fill={rgb,255: red,216; green,248; blue,216}, draw=black, shape=rectangle, inner sep=0.6mm, minimum height=1.5mm, minimum width=1.5mm]
\tikzstyle{Xb}=[fill={rgb,255: red,232; green,165; blue,165}, draw=black, shape=rectangle, inner sep=0.6mm, minimum height=1.5mm, minimum width=1.5mm]
\tikzstyle{Yb}=[fill={rgb,255: red,165; green,200; blue,232}, draw=black, shape=rectangle, inner sep=0.6mm, minimum height=1.5mm, minimum width=1.5mm]
\tikzstyle{Tb}=[fill={rgb,255: red,198; green,165; blue,232}, draw=black, shape=rectangle, inner sep=0.6mm, minimum height=1.5mm, minimum width=1.5mm]
\tikzstyle{Zt}=[fill={rgb,255: red,216; green,248; blue,216}, draw=black, shape=isosceles triangle, inner sep=0.6mm, minimum height=1.5mm, minimum width=1.5mm]
\tikzstyle{Xt}=[fill={rgb,255: red,232; green,165; blue,165}, draw=black, shape=isosceles triangle, inner sep=0.6mm, minimum height=1.5mm, minimum width=1.5mm]
\tikzstyle{Yt}=[fill={rgb,255: red,165; green,200; blue,232}, draw=black, shape=isosceles triangle, inner sep=0.6mm, minimum height=1.5mm, minimum width=1.5mm]
\tikzstyle{Tt}=[fill={rgb,255: red,198; green,165; blue,232}, draw=black, shape=isosceles triangle, inner sep=0.6mm, minimum height=1.5mm, minimum width=1.5mm]
\tikzstyle{Ztf}=[fill={rgb,255: red,216; green,248; blue,216}, draw=black, shape=isosceles triangle, rotate=180, inner sep=0.6mm, minimum height=1.5mm, minimum width=1.5mm]
\tikzstyle{Xtf}=[fill={rgb,255: red,232; green,165; blue,165}, draw=black, shape=isosceles triangle, rotate=180, inner sep=0.6mm, minimum height=1.5mm, minimum width=1.5mm]
\tikzstyle{Ytf}=[fill={rgb,255: red,165; green,200; blue,232}, draw=black, shape=isosceles triangle, rotate=180, inner sep=0.6mm, minimum height=1.5mm, minimum width=1.5mm]
\tikzstyle{Ttf}=[fill={rgb,255: red,198; green,165; blue,232}, draw=black, shape=isosceles triangle, rotate=180, inner sep=0.6mm, minimum height=1.5mm, minimum width=1.5mm]
\tikzstyle{Ztse}=[fill={rgb,255: red,216; green,248; blue,216}, draw=black, shape=isosceles triangle, rotate=315, inner sep=0.6mm, minimum height=1.5mm, minimum width=1.5mm]
\tikzstyle{Xtse}=[fill={rgb,255: red,232; green,165; blue,165}, draw=black, shape=isosceles triangle, rotate=315, inner sep=0.6mm, minimum height=1.5mm, minimum width=1.5mm]
\tikzstyle{Ytse}=[fill={rgb,255: red,165; green,200; blue,232}, draw=black, shape=isosceles triangle, rotate=315, inner sep=0.6mm, minimum height=1.5mm, minimum width=1.5mm]
\tikzstyle{Ttse}=[fill={rgb,255: red,198; green,165; blue,232}, draw=black, shape=isosceles triangle, rotate=315, inner sep=0.6mm, minimum height=1.5mm, minimum width=1.5mm]
\tikzstyle{Ztsw}=[fill={rgb,255: red,216; green,248; blue,216}, draw=black, shape=isosceles triangle, rotate=225, inner sep=0.6mm, minimum height=1.5mm, minimum width=1.5mm]
\tikzstyle{Xtsw}=[fill={rgb,255: red,232; green,165; blue,165}, draw=black, shape=isosceles triangle, rotate=225, inner sep=0.6mm, minimum height=1.5mm, minimum width=1.5mm]
\tikzstyle{Ytsw}=[fill={rgb,255: red,165; green,200; blue,232}, draw=black, shape=isosceles triangle, rotate=225, inner sep=0.6mm, minimum height=1.5mm, minimum width=1.5mm]
\tikzstyle{Ttsw}=[fill={rgb,255: red,198; green,165; blue,232}, draw=black, shape=isosceles triangle, rotate=225, inner sep=0.6mm, minimum height=1.5mm, minimum width=1.5mm]
\tikzstyle{Ztnw}=[fill={rgb,255: red,216; green,248; blue,216}, draw=black, shape=isosceles triangle, rotate=135, inner sep=0.6mm, minimum height=1.5mm, minimum width=1.5mm]
\tikzstyle{Xtnw}=[fill={rgb,255: red,232; green,165; blue,165}, draw=black, shape=isosceles triangle, rotate=135, inner sep=0.6mm, minimum height=1.5mm, minimum width=1.5mm]
\tikzstyle{Ytnw}=[fill={rgb,255: red,165; green,200; blue,232}, draw=black, shape=isosceles triangle, rotate=135, inner sep=0.6mm, minimum height=1.5mm, minimum width=1.5mm]
\tikzstyle{Ttnw}=[fill={rgb,255: red,198; green,165; blue,232}, draw=black, shape=isosceles triangle, rotate=135, inner sep=0.6mm, minimum height=1.5mm, minimum width=1.5mm]
\tikzstyle{Ztne}=[fill={rgb,255: red,216; green,248; blue,216}, draw=black, shape=isosceles triangle, rotate=45, inner sep=0.6mm, minimum height=1.5mm, minimum width=1.5mm]
\tikzstyle{Xtne}=[fill={rgb,255: red,232; green,165; blue,165}, draw=black, shape=isosceles triangle, rotate=45, inner sep=0.6mm, minimum height=1.5mm, minimum width=1.5mm]
\tikzstyle{Ytne}=[fill={rgb,255: red,165; green,200; blue,232}, draw=black, shape=isosceles triangle, rotate=45, inner sep=0.6mm, minimum height=1.5mm, minimum width=1.5mm]
\tikzstyle{Ttne}=[fill={rgb,255: red,198; green,165; blue,232}, draw=black, shape=isosceles triangle, rotate=45, inner sep=0.6mm, minimum height=1.5mm, minimum width=1.5mm]

\tikzstyle{simple}=[-]
\tikzstyle{ha}=[-, decorate, decoration={markings, mark= at position 0.5 with {\arrow{stealth}}}]
\tikzstyle{arrow}=[-, decorate, decoration={markings, mark= at position 0.7 with {\arrow{stealth}}}]
\tikzstyle{harrow}=[-, decorate, decoration={markings, mark= at position 0.825 with {\arrow{stealth}}}]
\tikzstyle{worra}=[-, decorate, decoration={markings, mark= at position 0.4 with {\arrow{stealth}}}]
\tikzstyle{hworra}=[-, decorate, decoration={markings, mark= at position 0.25 with {\arrow{stealth}}}]
\tikzstyle{ze}=[-, color={rgb,255: red,216; green,248; blue,216}]
\tikzstyle{xe}=[-, color={rgb,255: red,232; green,165; blue,165}]
\tikzstyle{ye}=[-, color={rgb,255: red,165; green,200; blue,232}]
\tikzstyle{te}=[-, color={rgb,255: red,198; green,165; blue,232}]
\tikzstyle{hadamard edge}=[-, color=blue, dashed, dash pattern=on 2pt off 0.7pt]
\tikzstyle{empty}=[-, color=black, dashed, dash pattern=on 1pt off 0.5pt]
\tikzstyle{brace edge}=[-, tikzit draw=blue, decorate, decoration={brace,amplitude=1mm,raise=-1mm}]
\tikzstyle{zh}=[-, decorate, decoration={markings, mark= at position 0.5 with {\arrow{stealth}}}]

\newtheorem{theorem}{Theorem}

\newtheorem{proposition}[theorem]{Proposition}

\theoremstyle{definition}
\newtheorem{definition}[theorem]{Definition}

\DeclareMathSymbol{\shortminus}{\mathbin}{AMSa}{"39}

\newcommand{\odal}[0]{\textarc{o}}
\newcommand{\algiz}[0]{\textarc{z}}

\newcommand{\kaun}[0]{\textarc{K}}

\newcommand{\letin}[3]{\mathrm{let}\; #1 = #2 \;\mathrm{in}\; #3}

\title{The Zeta Calculus}
\author{%
Nicklas Botö \qquad\qquad Fabian Forslund \\
\vspace{-.5cm}
\email{boton@chalmers.se \qquad\qquad fabfors@chalmers.se} \\
\institute{Chalmers University of Technology \\
Gothenburg, Sweden}
}
\def\titlerunning{The Zeta Calculus}
\def\authorrunning{N. Botö, F. Forslund}
\begin{document}
\maketitle

\begin{abstract}
We propose a quantum programming language that generalizes the $\lambda$-calculus. The language is non-linear; duplicated variables denote, not cloning of quantum data, but \emph{sharing} a qubit's state; that is, producing an entangled pair of qubits whose amplitudes are identical with respect to a chosen basis. The language has two abstraction operators, $\zeta$ and $\xi$, corresponding to the Z- and X-bases; each abstraction operator is also parameterised by a phase, indicating a rotation that is applied to the input before it is shared. We give semantics for the language in the ZX-calculus and prove its equational theory sound. We show how this language can provide a good representation of higher-order functions in the quantum world.
\end{abstract}

\section{Introduction}

Thanks to the \emph{no-cloning theorem} \cite{Wootters1982ACloned}, which famously states that a quantum state cannot be duplicated, most attempts to create a $\lambda$-calculus-like language for quantum computing have employed a \emph{linear} type system. For example, the systems Quipper\cite{Green2013Quipper}, Lineal\cite{Arrighi2017Lineal:Lambda-Calculus} and QWIRE\cite{Paykin2017QWIRE:Circuits} all provide linear type systems, with non-linear rules for classical data.

However, cloning of data is not the only possible way to interpret a duplicated variable. We can also interpret a duplicated variable as indicating \emph{sharing} of a qubit. That is, if the variable $x$ holds a reference to a qubit in the state $\alpha \ket{0} + \beta \ket{1}$, then an expression which uses $x$ twice should be interpreted as an instruction to form an entangled pair of qubits in state $\alpha \ket{00} + \beta \ket{11}$. This is the interpretation of duplicated variables in QML\cite{Altenkirch2005ALanguage}, which however does not include support for higher-order functions.

In this paper we introduce the \emph{$\zeta$-calculus}, a non-linear typed system with support for higher-order functions in which duplicated variables are interpreted as sharing. We provide the syntax, typing rules, operational semantics mapping terms in $\zeta$ to diagrams of the ZX-calculus\cite{Coecke2011InteractingDiagrammatics}\cite{vandeWetering2020ZX-calculusScientist} and an equational theory which is sound with respect to the rules of ZX. We show how the linear $\lambda$-calculus may be embedded in the $\zeta$-calculus, and also discuss a number of examples showing the capabilities of the language, including examples of terms demonstrating the use of higher-order functions.

Note that, in order to share a qubit, we must choose a basis. We call this operation \emph{sharing across the basis $\beta$}. Thus, sharing across the basis $Z = \{ \ket{0}, \ket{1} \}$ is the operation that maps $\alpha \ket{0} + \beta \ket{1}$ to $\alpha \ket{00} + \beta \ket{11}$, while sharing across the basis $X = \{ \ket{+}, \ket{-} \}$ is the operation that maps $\alpha \ket{+} + \beta \ket{-}$ to $\alpha \ket{++} + \beta \ket{--}$.

Instead of privileging the standard basis, the $\zeta$-calculus introduces a different binding abstraction for each basis. It has two binders, $\zeta$ and $\xi$, corresponding to the Z- and the X-bases. Thus, $\xi x M$ should be read as 'Perform the computation $M$ and, if the variable $x$ occurs more than once, then share the state of $x$ across the basis $X$', and $\zeta x M$ is read the same with respect to the Z-basis.

\subsection{Example: Higher-order sharing.} \label{subsec:hos}

\begin{figure}[h]
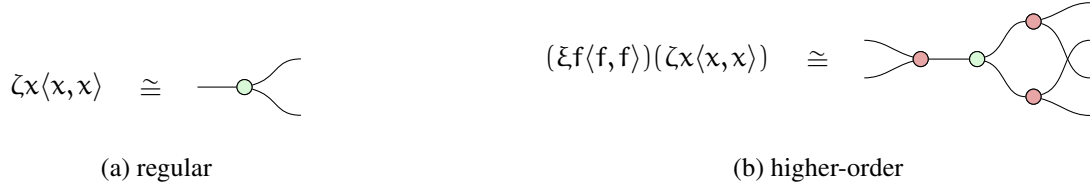
 
\centering
\begin{subfigure}[b]{.4\textwidth} 
\centering
\ctikzfig{tikz/single_sharing_example}
\caption{regular}
\label{fig:sharing_examples_regular}
\end{subfigure}
\hfill
\begin{subfigure}[b]{.5\textwidth} 
\centering
\ctikzfig{tikz/sharing_example}
\caption{higher-order}
\label{fig:sharing_examples_higher_order}
\end{subfigure}
\caption{Examples of sharing.}
\label{fig:sharing_examples}
\end{figure}

We wish to highlight two of the main features of the language's semantics: sharing and higher-order functions. In the $\zeta$-calculus, sharing in the basis $Z$ is represented by the term $\zeta x \langle x , x \rangle$, duplicating a variable introduced in $Z$. This should be read like the term $\lambda x . (x,x)$ in the $\lambda$-calculus. The semantics of this term, as a ZX-diagram, is presented in figure \ref{fig:sharing_examples_regular}.

The term \textit{higher-order function} (HOFs) should be familiar to the functional programmer, where a function accepts another function as input. In figure \ref{fig:sharing_examples_higher_order}, we apply the function $\xi f \langle f , f \rangle$ to the sharing function $\zeta x \langle x, x \rangle$. That is, we share the sharing function. Its derivation is in appendix \ref{appendix:sharing_sharing}.

\section{Syntax}

We define the set of terms $\algiz$ (\textit{algiz}) of the $\zeta$-calculus as shown in figure \ref{ir:syntax}. 
The variable $x$ is bound within the term $\beta^\alpha x M$, and we identify terms up to $\alpha$-conversion. We write $\beta x M$ for $\beta^0 x M$.

\begin{figure}[h] 
\begin{mathpar}
\inferrule*[right=unit]
    { }
    { * \in \algiz }

\inferrule*[right=var]
    { }
    { x \in \algiz }

\inferrule*[right=gen]
    { \beta \in \{ \zeta, \xi \} \\ 
    \alpha \in [0, 2 \pi) \\
    n \in \mathbb{Z} }
    { \beta^\alpha_n \in \algiz }

\inferrule*[right=abs$_\beta$]
    { \beta \in \{ \zeta, \xi \} \\
    \alpha \in [0, 2 \pi)
    \\ M \in \algiz }
    { \beta^\alpha \, x M \in \algiz }

\inferrule*[right=app]
    { M \in \algiz \\ N \in \algiz }
    { M N \in \algiz }

\inferrule*[right=tup]
    { M \in \algiz \\ N \in \algiz }
    { \langle M, N \rangle \in \algiz }

\inferrule*[right=let]
    { M \in \algiz \\ N \in \algiz \\ \beta \in \{\zeta, \xi\} } 
    { \letin{\langle x, y \rangle}{_\beta M}{N} \in \algiz }
\end{mathpar}
\caption{Syntax of the calculus.}
\label{ir:syntax}
\end{figure}

Informally, the abstraction $\beta^\theta x M$ can be read as 'Perform the computation $M$ on the variable $x$ rotated about the basis $\beta$ by an angle $\theta$, and if $x$ appears more than once in $M$ then it is shared into $M$. 
The rule \textsc{gen} generates values from the basis vectors of a given basis. For some basis $\beta$, angle $\alpha$ and positive $n$, states of the form $\ket{\beta_0}^{\otimes n} + e^{i \alpha}\ket{\beta_1}^{\otimes n}$ are introduced. For negative $n$ we get effects and for $n = 0$ the value becomes a scalar.
We define some commonly used syntactic sugar, 
\[
\hat{\beta}^\alpha \;:\equiv\; \beta^\alpha x x
\qquad
M \circ N \;:\equiv\; \beta^0 x M (N x)
\qquad
H \;:\equiv\; \hat{\zeta}^\frac{\pi}{2} \circ \hat{\xi}^\frac{\pi}{2} \circ \hat{\zeta}^\frac{\pi}{2}
\quad.
\]

\section{Typing}

The set of types is defined in figure \ref{ir:types}, consisting of numeral types, tensor products, and type duals.

\begin{figure}[ht] 
\begin{mathpar}
\inferrule
    { n \in \mathbb{N} }
    { \underline{n} \in Type }

\inferrule
    { A \in Type \\ B \in Type }
    { A \otimes B \in Type }

\inferrule
    { A \in Type }
    { A^* \in Type }
\end{mathpar}
\caption{Definition of types.}
\label{ir:types}
\end{figure}

A term of type $\underline{n}$ represents (a process that outputs) $n$ qubits.
Tensor products and dual types have their usual interpretation. 
We define the \emph{function type} $A \rightarrow B$ to be $A^* \otimes B$.
Since the category of Hilbert spaces is closed monoidal, a function $A \rightarrow B$ can be represented as a state of type $A^* \otimes B$ \cite{Baez2009PhysicsStone}.
The unit type $\top$ is defined as the zero numeral $\top := \underline{0}$. 
The typing contexts are given by the grammar $\Gamma ::= \emptyset \;|\; \Gamma, x :_\beta\! A$.

A judgement of the $\zeta$-calculus has the form $x_1 :_{\beta_1} A_1, \ldots, x_n :_{\beta_n} A_n \vdash M : B$. This denotes a quantum process that takes inputs of type $A_1$, \ldots, $A_n$ and outputs a state of type $B$. If the variable $x_i$ occurs more than once in the term $M$, then it is to be shared across basis $\beta_i$.

The typing rules of the $\zeta$-calculus are presented in figures \ref{ir:structure} and \ref{ir:typing}.

\begin{figure}[h]
\begin{mathpar}
\inferrule*[right=w]
    { \Gamma \vdash M : B  }
    { \Gamma, x :_\beta\! A \vdash M : B }

\inferrule*[right=c]
    { \Gamma, x_1 :_\beta\! A, \dots, x_n :_\beta\! A \vdash M : B  }
    { \Gamma, x :_\beta\! A \vdash M[x_1 := x, \dots, x_n := x] : B }
 
\inferrule*[right=x]
    {\Gamma_1, \Delta, \Gamma_2, \Phi, \Gamma_3 \vdash M : A}
    {\Gamma_1, \Phi, \Gamma_2, \Delta, \Gamma_3 \vdash M : A}
\end{mathpar}
\caption{Structural rules.}
\label{ir:structure}
\end{figure}

\begin{figure}[h] 
\begin{mathpar}
\inferrule*[right=u]
    { }
    { \Gamma \vdash * : \top }
    
\inferrule*[right=v]
    { x :_\beta\! A \in \Gamma }
    { \Gamma \vdash x : A }
    
\inferrule*[right=g]
    { }
    { \Gamma \vdash \beta_n^\alpha : \underline{n} }

\inferrule*[right=d]
    { }
    { \Gamma \vdash \beta_{-n}^\alpha : \underline{n} \to \top }

\inferrule*[right=b]
    { \Gamma, x :_\beta\! A \vdash M : B }
    { \Gamma \vdash \beta^\alpha x M : A \to  B }

\inferrule*[right=a]
    { \Gamma \vdash M : A \to B \\ \Gamma \vdash N : A }
    { \Gamma \vdash M N : B }

\inferrule*[right=t]
    { \Gamma \vdash M : A \\ \Gamma \vdash N : B }
    { \Gamma \vdash \langle M , N \rangle : A \otimes B }

\inferrule*[right=e]
    { \Gamma \vdash M : A \otimes B \\ \Gamma, x :_\beta\! A, y :_\beta\! B \vdash N : C }
    { \Gamma \vdash \letin{\langle x, y \rangle}{_\beta M}{N} : C }
\end{mathpar}
\caption{Typing rules.}
\label{ir:typing}
\end{figure}

Beyond the trivial axioms of (\textsc{u}) and (\textsc{v}) we have the type of the basis generators (\textsc{g}, \textsc{d}). In the case of $n \geq 0$ the type is simply a numeral equal to the size of the generator, while the negative case defines an effect (that is, a function into the unit type). 
The rule (\textsc{b}) introduces the function type.
The elimination rule of the function type is (\textsc{a}), the usual application rule.
The tensor product type is introduced by (\textsc{t}) and eliminated by the let-rule (\textsc{e}).
The structural rules for the system are weakening (\textsc{w}), contraction (\textsc{c}) and exchange (\textsc{x}).
\section{Semantics}

We give semantics for the system by mapping every derivable judgement to a diagram of the ZX-calculus as follows.
Every \emph{type} $A$ and \emph{context} $\Gamma$ is interpreted as a set of \emph{labels}, $\llbracket \Gamma \rrbracket$ and $\llbracket A \rrbracket$. The intention is that a derivable judgement $\Gamma \vdash M : A$ will be mapped to a diagram whose open input wires (left side of the diagram) are labelled by the elements of $\llbracket \Gamma \rrbracket$ and whose open output wires (right side of the diagram) are labelled by the elements of $\llbracket A \rrbracket$. One such labelling is presented below.
\begin{align*}
    \llbracket \underline{n} \rrbracket & := \{ 0, 1, \ldots, n-1 \} \\
    \llbracket A \otimes B \rrbracket & := \{ (0, a) : a \in \llbracket A \rrbracket \} \cup \{ (1, b) : b \in \llbracket B \rrbracket \} \\
    \llbracket A^* \rrbracket & := \{ a^* : a \in \llbracket A \rrbracket \}
\end{align*}

\begin{figure}[h]
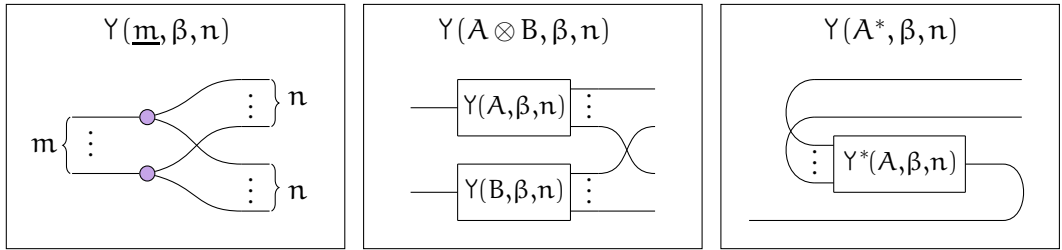

\ctikzfig{tikz/sharing_def}
\caption{Definition of the sharing operation $\kaun$ (\textit{kaun}).}
\label{fig:sharing_def}
\end{figure}

Define the diagrams $\kaun(A, \beta, n)$ for $A$ a type, $\beta \in \{ \zeta, \xi \}$ and $n \in \mathbb{N}$ as in figure \ref{fig:sharing_def}, where $\kaun^*$ is the horizontal reflection of the sharing diagram. The intention is that it denotes the operation that produces $n$ shared copies of a value of type $A$, shared across the basis $\beta$.

\begin{figure}[h]
\resizebox{\textwidth}{!}{\begin{tikzpicture}
	\begin{pgfonlayer}{nodelayer}
		\node [style=none] (0) at (7.25, -1) {};
		\node [style=none] (1) at (5.75, 0) {};
		\node [style=none] (2) at (5.75, -2) {};
		\node [style=none] (3) at (5.75, 0.75) {};
		\node [style=none] (4) at (5.75, -2.75) {};
		\node [style=none] (5) at (9.25, -1) {};
		\node [style=none] (6) at (3, 0.75) {};
		\node [style=none] (7) at (5.25, 0) {};
		\node [style=none] (8) at (5.25, -2) {};
		\node [style=none] (9) at (3, -2.75) {};
		\node [style=none] (10) at (3.75, 0) {};
		\node [style=none] (11) at (3.75, -2) {};
		\node [style=none] (12) at (3, 0) {};
		\node [style=none] (13) at (3, -2) {};
		\node [style=none] (14) at (3, -1) {};
		\node [style=none] (15) at (5.75, -1) {};
		\node [style=none] (16) at (2.5, 0.75) {$\Gamma_1$};
		\node [style=none] (17) at (2.5, 0) {$\Theta$};
		\node [style=none] (18) at (2.5, -1) {$\Gamma_2$};
		\node [style=none] (19) at (2.5, -2) {$\Delta$};
		\node [style=none] (20) at (2.5, -2.75) {$\Gamma_3$};
		\node [style=none] (21) at (9.75, -1) {$A$};
		\node [style=none] (22) at (6.5, -1) {$M$};
		\node [style=none] (23) at (5.75, 1) {};
		\node [style=none] (24) at (7.25, 1) {};
		\node [style=none] (25) at (7.25, -3) {};
		\node [style=none] (26) at (5.75, -3) {};
		\node [style=none] (27) at (6, 3) {$\inferrule*[right=x]
    {\Gamma_1, \Delta, \Gamma_2, \Theta, \Gamma_3 \vdash M : A}
    {\Gamma_1, \Theta, \Gamma_2, \Delta, \Gamma_3 \vdash M : A}$};
		\node [style=none] (28) at (-7, 2.5) {$\inferrule*[right=c]{ \Gamma,\; x_1 :_\beta\! A, \dots, x_n :_\beta\! A \; \vdash M : B}{\Gamma, x :_\beta\! A \vdash M[x_1 := x, \dots, x_n := x] : B}$};
		\node [style=none] (29) at (-14, 4.5) {};
		\node [style=none] (30) at (-0.25, 4.5) {};
		\node [style=none] (31) at (-0.25, -3.75) {};
		\node [style=none] (32) at (-14, -3.75) {};
		\node [style=none] (33) at (0.25, 4.5) {};
		\node [style=none] (34) at (11.75, 4.5) {};
		\node [style=none] (35) at (11.75, -3.75) {};
		\node [style=none] (36) at (0.25, -3.75) {};
		\node [style=none] (37) at (-4.5, -1.25) {};
		\node [style=none] (38) at (-6, -2.5) {};
		\node [style=none] (39) at (-6, -0.25) {};
		\node [style=none] (40) at (-2.5, -1.25) {};
		\node [style=none] (41) at (-7, -2.5) {};
		\node [style=none] (42) at (-10.75, -0.25) {};
		\node [style=none] (43) at (-7, -1.5) {};
		\node [style=none] (44) at (-6, -1.5) {};
		\node [style=none] (45) at (-11.25, -0.25) {$\Gamma$};
		\node [style=none] (46) at (-2, -1.25) {$B$};
		\node [style=none] (47) at (-5.25, -1.5) {$M$};
		\node [style=none] (48) at (-6, -3) {};
		\node [style=none] (49) at (-4.5, -3) {};
		\node [style=none] (50) at (-4.5, 0) {};
		\node [style=none] (51) at (-6, 0) {};
		\node [style=none] (52) at (-7, -2.75) {};
		\node [style=none] (53) at (-7, -1.25) {};
		\node [style=none] (54) at (-10, -1.25) {};
		\node [style=none] (55) at (-10, -2.75) {};
		\node [style=none] (56) at (-8.5, -2) {{\small $\kaun\!(A,\! \beta,\! n)$}};
		\node [style=none] (57) at (-6.5, -1.75) {$\vdots$};
		\node [style=none] (58) at (-10.75, -2) {};
		\node [style=none] (59) at (-12, -2) {$x :_\beta\! A$};
		\node [style=none] (60) at (-10, -2) {};
		\node [style=none] (61) at (-19.25, 2.5) {$\inferrule*[right=w]{ \Gamma \vdash M : B }{ \Gamma, x :_\beta\! A \vdash M : B}$};
		\node [style=none] (62) at (-24.25, 4.5) {};
		\node [style=none] (63) at (-14.5, 4.5) {};
		\node [style=none] (64) at (-14.5, -3.75) {};
		\node [style=none] (65) at (-24.25, -3.75) {};
		\node [style=none] (69) at (-17, -0.5) {};
		\node [style=none] (71) at (-21.5, -0.5) {};
		\node [style=none] (74) at (-22, -0.5) {$\Gamma$};
		\node [style=none] (75) at (-16.5, -0.5) {$A$};
		\node [style=box] (90) at (-19, -0.5) {$M$};
		\node [style=none] (91) at (-17, -2.75) {};
		\node [style=none] (92) at (-17, -1.75) {};
		\node [style=none] (93) at (-17, -3) {};
		\node [style=none] (94) at (-17, -1.5) {};
		\node [style=none] (95) at (-20, -1.5) {};
		\node [style=none] (96) at (-20, -3) {};
		\node [style=none] (97) at (-18.5, -2.25) {{\small $\kaun\!(A,\! \beta,\! 0)$}};
		\node [style=none] (98) at (-21.5, -2.25) {};
		\node [style=none] (99) at (-22.75, -2.25) {$x :_\beta\! A$};
		\node [style=none] (100) at (-20, -2.25) {};
	\end{pgfonlayer}
	\begin{pgfonlayer}{edgelayer}
		\draw [style=simple] (6.center) to (3.center);
		\draw [style=simple] (7.center) to (1.center);
		\draw [style=simple] (8.center) to (2.center);
		\draw [style=simple] (9.center) to (4.center);
		\draw [style=simple] (0.center) to (5.center);
		\draw [style=simple] (12.center) to (10.center);
		\draw [style=simple] (13.center) to (11.center);
		\draw [style=simple, in=180, out=0] (10.center) to (8.center);
		\draw [style=simple, in=-180, out=0] (11.center) to (7.center);
		\draw [style=simple] (14.center) to (15.center);
		\draw [style=simple] (25.center) to (26.center);
		\draw [style=simple] (26.center) to (23.center);
		\draw [style=simple] (23.center) to (24.center);
		\draw [style=simple] (24.center) to (25.center);
		\draw (32.center) to (31.center);
		\draw (31.center) to (30.center);
		\draw [in=0, out=180] (30.center) to (29.center);
		\draw (29.center) to (32.center);
		\draw (36.center) to (35.center);
		\draw (35.center) to (34.center);
		\draw [in=0, out=180] (34.center) to (33.center);
		\draw (33.center) to (36.center);
		\draw [style=simple] (41.center) to (38.center);
		\draw [style=simple] (42.center) to (39.center);
		\draw [style=simple] (37.center) to (40.center);
		\draw [style=simple] (43.center) to (44.center);
		\draw [style=simple] (50.center) to (51.center);
		\draw [style=simple] (51.center) to (48.center);
		\draw [style=simple] (48.center) to (49.center);
		\draw [style=simple] (49.center) to (50.center);
		\draw (55.center) to (52.center);
		\draw (52.center) to (53.center);
		\draw (53.center) to (54.center);
		\draw (54.center) to (55.center);
		\draw (58.center) to (60.center);
		\draw (65.center) to (64.center);
		\draw (64.center) to (63.center);
		\draw [in=0, out=180] (63.center) to (62.center);
		\draw (62.center) to (65.center);
		\draw [style=simple] (71.center) to (90);
		\draw [style=simple] (90) to (69.center);
		\draw (96.center) to (93.center);
		\draw (93.center) to (94.center);
		\draw (94.center) to (95.center);
		\draw (95.center) to (96.center);
		\draw (98.center) to (100.center);
	\end{pgfonlayer}
\end{tikzpicture}}
\caption{Interpretation of structural rules.}
\label{fig:structural_rules}
\end{figure}
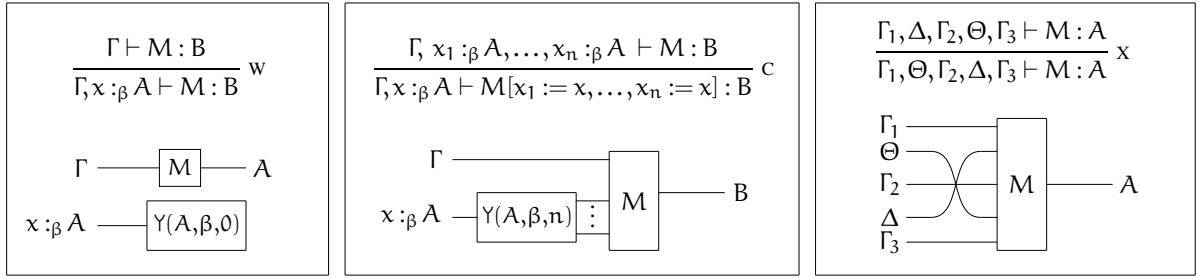

These operations can then be used to define the interpretation of the structural rules, see figure \ref{fig:structural_rules}.

Define the sharing of contexts $\kaun(\Gamma, n)$ as the operation that duplicates each variable in a context in its introduced basis ($x :_\beta\! A \mapsto \kaun(A, \beta, n)$) and swaps them accordingly into $n$ contexts. This is then used to produce multiple copies of $\Gamma$ in the operational semantics. Note that every wire and spider multiplies over the size, that is the number of labels, of its type. Similar to how spiders and wires act for integers in the SZX-calculus \cite{Carette2019SZX-calculus:Reasoning}.

\begin{figure}[ht] 
\resizebox{\textwidth}{!}{\begin{tikzpicture}
	\begin{pgfonlayer}{nodelayer}
		\node [style=none] (0) at (-11, 9.25) {$\inferrule*[right=u]{ }{\Gamma \vdash * : \top }$};
		\node [style=none] (1) at (-14.5, 10.75) {};
		\node [style=none] (2) at (-7.75, 10.75) {};
		\node [style=none] (3) at (-7.75, 5.5) {};
		\node [style=none] (4) at (-14.5, 5.5) {};
		\node [style=none] (5) at (-11.75, 7.5) {};
		\node [style=none] (6) at (-10.75, 7.5) {};
		\node [style=none] (7) at (-10.75, 6.5) {};
		\node [style=none] (8) at (-11.75, 6.5) {};
		\node [style=none] (9) at (-3.75, 9.25) {$\inferrule*[right=v]{  x :_\beta\! A \in \Gamma }{ \Gamma \vdash x : A}$};
		\node [style=none] (10) at (-2, 6.75) {};
		\node [style=none] (11) at (-1.5, 6.75) {$A$};
		\node [style=none] (14) at (-7.25, 10.75) {};
		\node [style=none] (15) at (-0.25, 10.75) {};
		\node [style=none] (16) at (-0.25, 5.5) {};
		\node [style=none] (17) at (-7.25, 5.5) {};
		\node [style=none] (18) at (3.75, 9.25) {$\inferrule*[right=g]{ }{\Gamma \vdash \beta_n^\alpha : \underline{n}}$};
		\node [style=Tp] (19) at (2.75, 7) {$\alpha$};
		\node [style=none] (20) at (4, 7.75) {};
		\node [style=none] (21) at (4.75, 7) {$\underline{n}$};
		\node [style=none] (50) at (-7.25, 3.25) {$\inferrule*[right=b]{\Gamma, x :_\beta\! A \; \vdash M : B}{\Gamma \vdash \beta^\alpha x M : A \to B}$};
		\node [style=Tp] (51) at (-6.75, 0.25) {$\alpha$};
		\node [style=none] (52) at (-4.5, 0.25) {};
		\node [style=none] (53) at (-4.5, -1.75) {};
		\node [style=none] (54) at (-3.75, 0.25) {$A^*$};
		\node [style=none] (55) at (-3.75, -1.75) {$B$};
		\node [style=none] (56) at (-7.25, -1.5) {};
		\node [style=none] (57) at (-7.25, -2) {};
		\node [style=none] (58) at (-11.5, -2) {};
		\node [style=none] (59) at (-12.25, -2) {$\Gamma$};
		\node [style=box] (60) at (-6.75, -1.75) {$M$};
		\node [style=none] (61) at (-7.25, 0.25) {};
		\node [style=none] (62) at (-10, 0.25) {};
		\node [style=none] (63) at (-10, -1.5) {};
		\node [style=none] (64) at (-9, -1) {$x :_\beta\! A$};
		\node [style=none] (79) at (7.5, 3) {$\inferrule*[right=a]{\Gamma \vdash M : A \to B \\ \Gamma \vdash N : A}{\Gamma \vdash M N : B}$};
		\node [style=none] (80) at (10.5, -2) {};
		\node [style=none] (81) at (8.75, 0) {};
		\node [style=none] (82) at (11.25, -2) {$B$};
		\node [style=none] (83) at (4.5, 0) {};
		\node [style=none] (85) at (4, 0) {$\Gamma$};
		\node [style=box] (86) at (6.5, 0) {$N$};
		\node [style=none] (87) at (4.5, -1.75) {};
		\node [style=box] (88) at (6.5, -1.75) {$M$};
		\node [style=none] (91) at (7, -2) {};
		\node [style=none] (92) at (7, 0) {};
		\node [style=none] (93) at (8.75, -1.5) {};
		\node [style=none] (94) at (7, -1.5) {};
		\node [style=none] (95) at (8, 0.5) {$A$};
		\node [style=none] (96) at (8, -1) {$A^*$};
		\node [style=none] (101) at (-7, -5.75) {$\inferrule*[right=t]{\Gamma \vdash M : A \\ \Gamma \vdash N : B}{\Gamma \ \vdash \langle M, N \rangle : A \otimes B}$};
		\node [style=none] (102) at (-5.25, -9) {};
		\node [style=none] (103) at (-5.25, -10.5) {};
		\node [style=none] (104) at (-4.5, -9) {$A$};
		\node [style=none] (105) at (-4.5, -10.5) {$B$};
		\node [style=none] (106) at (-8.25, -10.5) {};
		\node [style=none] (107) at (-10, -10.5) {};
		\node [style=box] (109) at (-7.75, -10.5) {$N$};
		\node [style=none] (110) at (-8.25, -9) {};
		\node [style=box] (111) at (-7.75, -9) {$M$};
		\node [style=none] (112) at (-10, -9) {};
		\node [style=none] (114) at (-7.25, -9) {};
		\node [style=none] (115) at (-7.25, -10.5) {};
		\node [style=none] (116) at (7.5, -5.5) {$\inferrule*[right=e]
    { \Gamma \vdash M : A \otimes B \\ \Gamma, x :_\beta\! A, y :_\beta\! B \vdash N : C }
    { \Gamma \vdash \letin{\langle x, y \rangle}{_\beta M}{N} : C }$};
		\node [style=none] (117) at (0.25, -3.75) {};
		\node [style=none] (118) at (14.5, -3.75) {};
		\node [style=none] (119) at (14.5, -12) {};
		\node [style=none] (120) at (0.25, -12) {};
		\node [style=box] (121) at (6.5, -9) {$M$};
		\node [style=mbox] (122) at (8.75, -9.75) {$N$};
		\node [style=none] (123) at (8.25, -8.75) {};
		\node [style=none] (124) at (8.25, -10.5) {};
		\node [style=none] (125) at (8.25, -9.25) {};
		\node [style=none] (126) at (7, -8.75) {};
		\node [style=none] (127) at (7, -9.25) {};
		\node [style=none] (129) at (6.25, -10.5) {};
		\node [style=none] (132) at (10.75, -9.75) {};
		\node [style=none] (133) at (11.25, -9.75) {$C$};
		\node [style=none] (138) at (0.25, 5) {};
		\node [style=none] (139) at (14.5, 5) {};
		\node [style=none] (140) at (14.5, -3.25) {};
		\node [style=none] (141) at (0.25, -3.25) {};
		\node [style=none] (142) at (-14.5, -3.75) {};
		\node [style=none] (143) at (-0.25, -3.75) {};
		\node [style=none] (144) at (-0.25, -12) {};
		\node [style=none] (145) at (-14.5, -12) {};
		\node [style=none] (146) at (-14.5, 5) {};
		\node [style=none] (147) at (-0.25, 5) {};
		\node [style=none] (148) at (-0.25, -3.25) {};
		\node [style=none] (149) at (-14.5, -3.25) {};
		\node [style=none] (154) at (0.25, 10.75) {};
		\node [style=none] (155) at (7, 10.75) {};
		\node [style=none] (156) at (7, 5.5) {};
		\node [style=none] (157) at (0.25, 5.5) {};
		\node [style=none] (158) at (7.5, 10.75) {};
		\node [style=none] (159) at (14.5, 10.75) {};
		\node [style=none] (160) at (14.5, 5.5) {};
		\node [style=none] (161) at (7.5, 5.5) {};
		\node [style=none] (162) at (4, 6.25) {};
		\node [style=none] (163) at (4.25, 7.75) {};
		\node [style=none] (164) at (4.25, 6.25) {};
		\node [style=none] (165) at (3.75, 7.25) {$\vdots$};
		\node [style=none] (166) at (11, 9.25) {$\inferrule*[right=d]{ }{\Gamma \vdash \beta_{-n}^\alpha : \underline{n} \to \top}$};
		\node [style=none] (180) at (4.5, -9) {};
		\node [style=none] (182) at (4, -9) {$\Gamma$};
		\node [style=none] (183) at (4.5, -10.5) {};
		\node [style=none] (201) at (-4.5, 6.75) {};
		\node [style=none] (202) at (-5.75, 6.75) {$x :_\beta\! A$};
		\node [style=none] (204) at (-10.5, -9) {$\Gamma$};
		\node [style=Tp] (217) at (10, 7) {$\alpha$};
		\node [style=none] (218) at (11.25, 7.75) {};
		\node [style=none] (219) at (11.25, 6.25) {};
		\node [style=none] (220) at (11.5, 7.75) {};
		\node [style=none] (221) at (11.5, 6.25) {};
		\node [style=none] (222) at (11, 7.25) {$\vdots$};
		\node [style=none] (226) at (12, 7) {$\underline{n}^*$};
		\node [style=none] (227) at (-10.5, -10.5) {$\Gamma$};
		\node [style=none] (228) at (4, -1.75) {$\Gamma$};
		\node [style=none] (229) at (4, -10.5) {$\Gamma$};
	\end{pgfonlayer}
	\begin{pgfonlayer}{edgelayer}
		\draw (1.center) to (2.center);
		\draw (2.center) to (3.center);
		\draw (4.center) to (3.center);
		\draw (4.center) to (1.center);
		\draw [style=empty] (8.center)
			 to (5.center)
			 to (6.center)
			 to (7.center)
			 to cycle;
		\draw (15.center)
			 to [in=0, out=180] (14.center)
			 to (17.center)
			 to (16.center)
			 to cycle;
		\draw [in=-180, out=15, looseness=1.25] (19) to (20.center);
		\draw (51) to (52.center);
		\draw (57.center) to (58.center);
		\draw (60) to (53.center);
		\draw (61.center) to (51);
		\draw (62.center) to (61.center);
		\draw (63.center) to (56.center);
		\draw [bend left=90, looseness=1.75] (63.center) to (62.center);
		\draw (91.center) to (80.center);
		\draw (92.center) to (81.center);
		\draw [bend right=90, looseness=2.00] (93.center) to (81.center);
		\draw (94.center) to (93.center);
		\draw (106.center) to (107.center);
		\draw (112.center) to (110.center);
		\draw (114.center) to (102.center);
		\draw (115.center) to (103.center);
		\draw (120.center)
			 to (119.center)
			 to (118.center)
			 to [in=0, out=180] (117.center)
			 to cycle;
		\draw [style=simple] (126.center) to (123.center);
		\draw [style=simple] (127.center) to (125.center);
		\draw [style=simple] (124.center) to (129.center);
		\draw [style=simple] (122) to (132.center);
		\draw (141.center) to (140.center);
		\draw (140.center) to (139.center);
		\draw [in=0, out=180] (139.center) to (138.center);
		\draw (138.center) to (141.center);
		\draw (145.center) to (144.center);
		\draw (144.center) to (143.center);
		\draw [in=0, out=180] (143.center) to (142.center);
		\draw (142.center) to (145.center);
		\draw (149.center) to (148.center);
		\draw (148.center) to (147.center);
		\draw [in=0, out=180] (147.center) to (146.center);
		\draw (146.center) to (149.center);
		\draw (154.center) to (155.center);
		\draw (155.center) to (156.center);
		\draw (157.center) to (156.center);
		\draw (157.center) to (154.center);
		\draw (158.center) to (159.center);
		\draw (159.center) to (160.center);
		\draw (161.center) to (160.center);
		\draw (161.center) to (158.center);
		\draw [in=180, out=-15, looseness=1.25] (19) to (162.center);
		\draw [style=brace edge] (163.center) to (164.center);
		\draw (83.center) to (86);
		\draw (87.center) to (88);
		\draw (183.center) to (129.center);
		\draw (180.center) to (121);
		\draw [in=-180, out=15, looseness=1.25] (217) to (218.center);
		\draw [in=180, out=-15, looseness=1.25] (217) to (219.center);
		\draw [style=brace edge] (220.center) to (221.center);
		\draw [style=simple] (201.center) to (10.center);
	\end{pgfonlayer}
\end{tikzpicture}}
\caption{Operational semantics of the calculus.}
\label{fig:operational_semantics}
\end{figure}
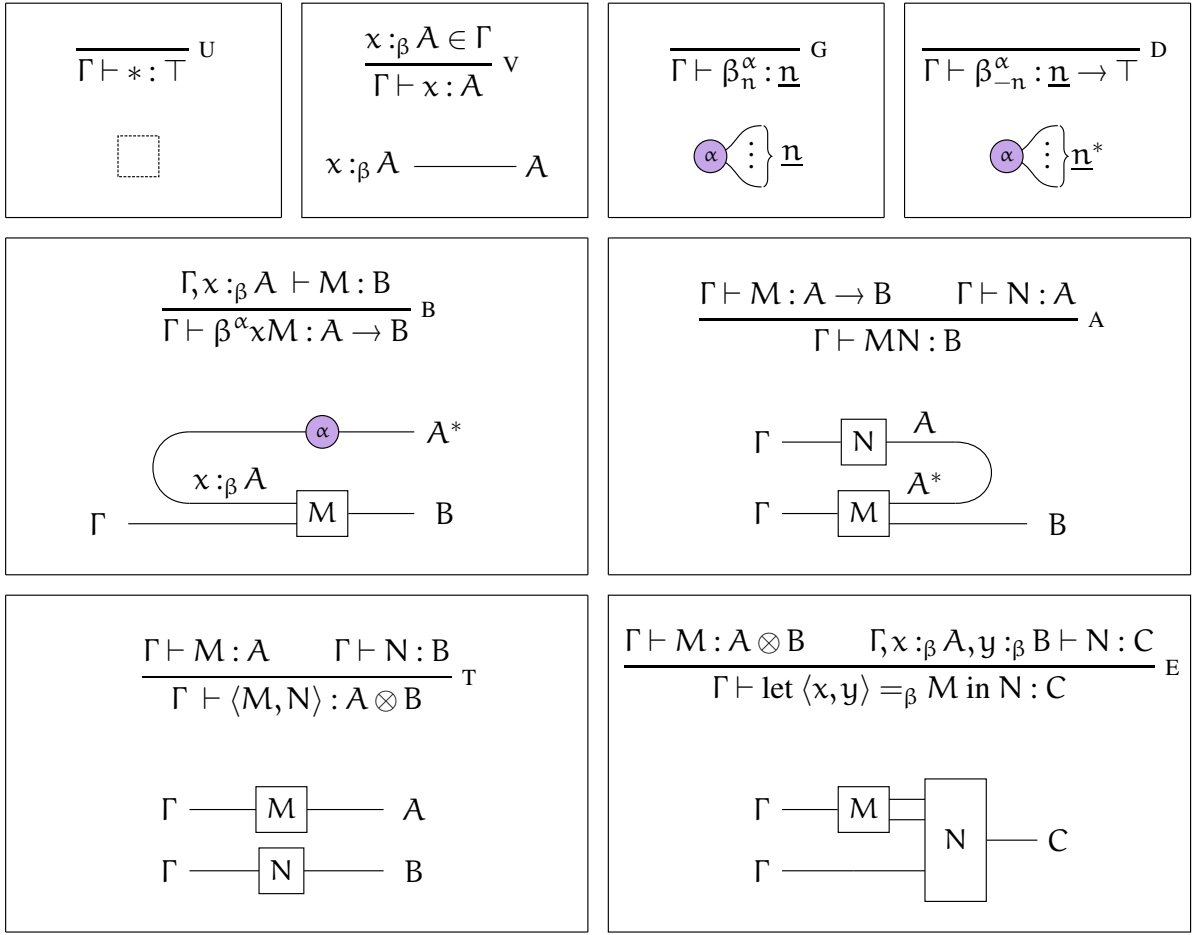

Now, we define the semantics of $\zeta$ as string diagram typing judgements in figure \ref{fig:operational_semantics}.

\newpage
We wish to define the familiar concept of substitution from the $\lambda$-calculus. This has some difficulties since not all terms can be substituted when sharing is involved. To aid in the proof of substitution we define the following. 

\begin{definition}\label{df:commshare}
Let $\Gamma \vdash M : A$ be a derivable judgement, and $\beta$ be a basis. Then we say $M$
\emph{commutes with sharing} over $\beta$ and $\Gamma$ iff
\ctikzfig{tikz/sharing_commutes}
\end{definition}

We will use the rules of the ZX-calculus in the equational theory of $\zeta$ to define more specifically what kinds of terms do commute over sharing. Using this definition, we can prove the following statement introducing substitution.

\newpage
\begin{proposition}[Substitution] \label{prop:subst}
Let $\Gamma, x :_\beta\! A, \Delta \vdash M : B$, and $\Theta \vdash N : A$. If $N$ commutes with sharing over $\beta$ and $\Theta$, then
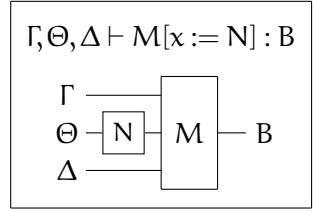
\ctikzfig{tikz/share_commute_N1}
\end{proposition}

\begin{proof}
The proof is by induction on the derivation of $\Gamma, x :_\beta\! A, \Delta \vdash M : B$. We give the details for the case of contraction here. Let $\Gamma, x_1 :_\beta\! A, \dots, x_n :_\beta\! A, \Delta \vdash M : B$, and $\Theta \vdash N : A$. Then, since $N$ commutes with sharing over $\beta$ and $\Theta$ we have

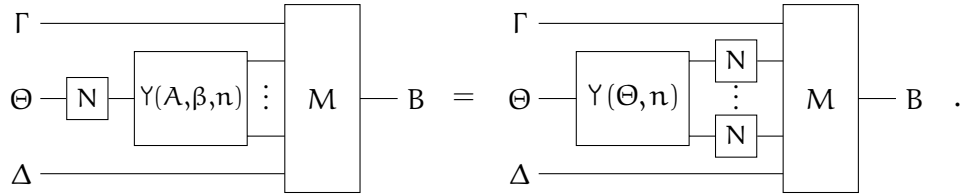
\[\begin{tikzpicture}
	\begin{pgfonlayer}{nodelayer}
		\node [style=none] (0) at (1.5, 2.5) {};
		\node [style=none] (1) at (3.5, 2.5) {};
		\node [style=none] (2) at (1.5, -2.5) {};
		\node [style=none] (3) at (3.5, -2.5) {};
		\node [style=none] (4) at (-5, 2) {};
		\node [style=none] (5) at (-0.25, 1) {};
		\node [style=none] (6) at (-0.25, -1) {};
		\node [style=none] (7) at (1.5, 2) {};
		\node [style=none] (8) at (1.5, 1) {};
		\node [style=none] (9) at (1.5, -1) {};
		\node [style=none] (10) at (-5, -2) {};
		\node [style=none] (11) at (-5, 0) {};
		\node [style=none] (12) at (1.5, -2) {};
		\node [style=none] (13) at (-1, 1) {};
		\node [style=none] (14) at (-1, -1) {};
		\node [style=none] (15) at (0.25, 0.25) {$\vdots$};
		\node [style=none] (16) at (-5.5, 0) {$\Theta$};
		\node [style=none] (17) at (-5.5, 2) {$\Gamma$};
		\node [style=none] (18) at (-5.5, -2) {$\Delta$};
		\node [style=none] (19) at (4.5, 0) {};
		\node [style=none] (20) at (3.5, 0) {};
		\node [style=none] (21) at (2.5, 0) {$M$};
		\node [style=none] (22) at (5, 0) {$B$};
		\node [style=box] (23) at (0.25, 1) {$N$};
		\node [style=box] (24) at (0.25, -1) {$N$};
		\node [style=none] (25) at (-1, 1) {};
		\node [style=none] (26) at (-1, -1) {};
		\node [style=none] (27) at (-1, 1.25) {};
		\node [style=none] (28) at (-1, -1.25) {};
		\node [style=none] (29) at (-4, -1.25) {};
		\node [style=none] (30) at (-4, 1.25) {};
		\node [style=none] (31) at (-2.5, 0) {$\kaun(\Theta,n)$};
		\node [style=none] (32) at (-4, 0) {};
		\node [style=none] (49) at (-11.75, 2.5) {};
		\node [style=none] (50) at (-9.75, 2.5) {};
		\node [style=none] (51) at (-11.75, -2.5) {};
		\node [style=none] (52) at (-9.75, -2.5) {};
		\node [style=none] (53) at (-18.25, 2) {};
		\node [style=none] (56) at (-11.75, 2) {};
		\node [style=none] (57) at (-11.75, 1) {};
		\node [style=none] (58) at (-11.75, -1) {};
		\node [style=none] (59) at (-18.25, -2) {};
		\node [style=none] (60) at (-18.25, 0) {};
		\node [style=none] (61) at (-11.75, -2) {};
		\node [style=none] (62) at (-12.75, 1) {};
		\node [style=none] (63) at (-12.75, -1) {};
		\node [style=none] (64) at (-12.25, 0.25) {$\vdots$};
		\node [style=none] (65) at (-18.75, 0) {$\Theta$};
		\node [style=none] (66) at (-18.75, 2) {$\Gamma$};
		\node [style=none] (67) at (-18.75, -2) {$\Delta$};
		\node [style=none] (68) at (-8.75, 0) {};
		\node [style=none] (69) at (-9.75, 0) {};
		\node [style=none] (70) at (-10.75, 0) {$M$};
		\node [style=none] (71) at (-8.25, 0) {$B$};
		\node [style=none] (74) at (-12.75, 1) {};
		\node [style=none] (75) at (-12.75, -1) {};
		\node [style=none] (76) at (-12.75, 1.25) {};
		\node [style=none] (77) at (-12.75, -1.25) {};
		\node [style=none] (78) at (-15.75, -1.25) {};
		\node [style=none] (79) at (-15.75, 1.25) {};
		\node [style=none] (80) at (-14.25, 0) {{\small $\kaun\!(A,\! \beta,\! n)$}};
		\node [style=none] (81) at (-15.75, 0) {};
		\node [style=box] (82) at (-17, 0) {$N$};
		\node [style=none] (83) at (-7, 0) {$=$};
	\end{pgfonlayer}
	\begin{pgfonlayer}{edgelayer}
		\draw (3.center)
			 to (1.center)
			 to (0.center)
			 to (2.center)
			 to cycle;
		\draw (4.center) to (7.center);
		\draw (12.center) to (10.center);
		\draw [in=-180, out=0, looseness=1.25] (13.center) to (5.center);
		\draw [in=180, out=0, looseness=1.25] (14.center) to (6.center);
		\draw (20.center) to (19.center);
		\draw (24) to (9.center);
		\draw (24) to (6.center);
		\draw (23) to (8.center);
		\draw (23) to (5.center);
		\draw (29.center)
			 to (30.center)
			 to (27.center)
			 to (28.center)
			 to cycle;
		\draw (11.center) to (32.center);
		\draw (25.center) to (13.center);
		\draw (14.center) to (26.center);
		\draw (52.center)
			 to (50.center)
			 to (49.center)
			 to (51.center)
			 to cycle;
		\draw (53.center) to (56.center);
		\draw (61.center) to (59.center);
		\draw (69.center) to (68.center);
		\draw (78.center)
			 to (79.center)
			 to (76.center)
			 to (77.center)
			 to cycle;
		\draw (74.center) to (62.center);
		\draw (63.center) to (75.center);
		\draw [style=simple] (74.center) to (57.center);
		\draw [style=simple] (75.center) to (58.center);
		\draw [style=simple] (60.center) to (82);
		\draw [style=simple] (82) to (81.center);
	\end{pgfonlayer}
\end{tikzpicture} \quad. \]

From which we obtain $\Gamma, \Theta, \Delta \vdash M[x := N] : B$.
\end{proof}

\section{Equational theory}

In this section we introduce some simple equational rules on the terms of the $\zeta$-calculus. We introduce a relation $M \equiv N$ to denote that two terms are equal. This relation under a predicate $\Phi$ is related to the semantics of $\zeta$ for all valid judgements on the terms in equation \eqref{eq:eqdef}.

\begin{equation} \label{eq:eqdef}
M \equiv N \quad (\Phi)
\qquad\implies\qquad
\inferrule{\Gamma \vdash M : A \\ \Gamma \vdash N : A \\ \Phi}{M \equiv N}
\end{equation}

Then, if two terms can be made equal by the reflexive transitive closure of ($\equiv$) we write $\zeta \vdash M \equiv N$. 

\subsection{\texorpdfstring{Embedding the linear $\lambda$-calculus}{Embedding the linear λ-calculus}}
The linear $\lambda$-calculus can be embedded in $\zeta$ by defining the $\lambda$-abstraction $\lambda x M :\equiv \beta^0 x M$, where $x$ occurs only once in $M$ (written as $\omega_x(M) = 1$). The identity removal rule of the ZX-calculus then allows us to remove the spider from the interpretation of the abstraction all-together, producing a diagram on the form in equation \eqref{eq:lambda_basis_def}.

\begin{equation} \label{eq:lambda_basis_def}
\begin{tikzpicture}
	\begin{pgfonlayer}{nodelayer}
		\node [style=none] (0) at (-3.5, 0.75) {};
		\node [style=none] (1) at (-3.5, -1) {};
		\node [style=none] (2) at (-2.75, 0.75) {$A^*$};
		\node [style=none] (3) at (-2.75, -1) {$B$};
		\node [style=none] (4) at (-5.5, -0.75) {};
		\node [style=none] (5) at (-5.5, -1.25) {};
		\node [style=none] (6) at (-7.5, -1.25) {};
		\node [style=none] (7) at (-8.25, -1.25) {$\Gamma$};
		\node [style=box] (8) at (-5, -1) {$M$};
		\node [style=none] (9) at (-5.5, 0.75) {};
		\node [style=none] (10) at (-6, 0.75) {};
		\node [style=none] (11) at (-6, -0.75) {};
		\node [style=T] (12) at (-5, 0.75) {};
		\node [style=none] (13) at (7, 0.75) {};
		\node [style=none] (14) at (7, -1) {};
		\node [style=none] (15) at (7.75, 0.75) {$A^*$};
		\node [style=none] (16) at (7.75, -1) {$B$};
		\node [style=none] (17) at (5, -0.75) {};
		\node [style=none] (18) at (5, -1.25) {};
		\node [style=none] (19) at (3, -1.25) {};
		\node [style=none] (20) at (2.25, -1.25) {$\Gamma$};
		\node [style=box] (21) at (5.5, -1) {$M$};
		\node [style=none] (22) at (5, 0.75) {};
		\node [style=none] (23) at (4.5, 0.75) {};
		\node [style=none] (24) at (4.5, -0.75) {};
		\node [style=none] (25) at (0, 0) {$\overset{(id)}{=}$};
	\end{pgfonlayer}
	\begin{pgfonlayer}{edgelayer}
		\draw (5.center) to (6.center);
		\draw (8) to (1.center);
		\draw (10.center) to (9.center);
		\draw (11.center) to (4.center);
		\draw [bend left=90, looseness=1.75] (11.center) to (10.center);
		\draw [style=simple] (9.center) to (12);
		\draw [style=simple] (12) to (0.center);
		\draw (18.center) to (19.center);
		\draw (21) to (14.center);
		\draw (23.center) to (22.center);
		\draw (24.center) to (17.center);
		\draw [bend left=90, looseness=1.75] (24.center) to (23.center);
		\draw [style=simple] (22.center) to (13.center);
	\end{pgfonlayer}
\end{tikzpicture}
\end{equation}

Then, since variables introduced by the $\lambda$-basis only occur once, no sharing occurs and every term substituted for it commutes through. By proposition \ref{prop:subst} we show that $\beta$-reduction holds for all terms when substituting a variable introduced in this way.

\begin{equation}
\begin{tikzpicture}
	\begin{pgfonlayer}{nodelayer}
		\node [style=none] (0) at (-3, 0.25) {};
		\node [style=none] (1) at (-2, -1) {};
		\node [style=none] (2) at (-1.25, -1) {$A$};
		\node [style=none] (3) at (-4, -0.75) {};
		\node [style=none] (4) at (-4, -1.25) {};
		\node [style=none] (5) at (-6, -1.25) {};
		\node [style=none] (6) at (-6.75, -1.25) {$\Gamma$};
		\node [style=box] (7) at (-3.5, -1) {$M$};
		\node [style=none] (8) at (-4, 0.25) {};
		\node [style=none] (9) at (-4.5, 0.25) {};
		\node [style=none] (10) at (-4.5, -0.75) {};
		\node [style=box] (11) at (-4, 1.25) {$N$};
		\node [style=none] (12) at (-3, 1.25) {};
		\node [style=none] (13) at (-3, 0.25) {};
		\node [style=none] (14) at (-3.5, 1.25) {};
		\node [style=none] (15) at (-6.75, 1.25) {$\Delta$};
		\node [style=none] (16) at (-4.5, 1.25) {};
		\node [style=none] (17) at (-6, 1.25) {};
		\node [style=none] (18) at (6.75, 0) {};
		\node [style=none] (19) at (7.5, 0) {$A$};
		\node [style=none] (20) at (4.75, 0.75) {};
		\node [style=none] (21) at (4.75, -0.75) {};
		\node [style=none] (22) at (2.5, -0.75) {};
		\node [style=none] (23) at (1.75, -0.75) {$\Gamma$};
		\node [style=box] (24) at (3.5, 0.75) {$N$};
		\node [style=none] (25) at (4, 0.75) {};
		\node [style=none] (26) at (1.75, 0.75) {$\Delta$};
		\node [style=none] (27) at (3.25, 0.75) {};
		\node [style=none] (28) at (2.5, 0.75) {};
		\node [style=mbox] (29) at (5.25, 0) {$M$};
		\node [style=none] (30) at (0, 0) {$=$};
	\end{pgfonlayer}
	\begin{pgfonlayer}{edgelayer}
		\draw (4.center) to (5.center);
		\draw (7) to (1.center);
		\draw (9.center) to (8.center);
		\draw (10.center) to (3.center);
		\draw [bend left=90, looseness=1.75] (10.center) to (9.center);
		\draw [style=simple] (8.center) to (0.center);
		\draw [bend right=90, looseness=1.75] (13.center) to (12.center);
		\draw [style=simple] (14.center) to (12.center);
		\draw [style=simple] (17.center) to (16.center);
		\draw (21.center) to (22.center);
		\draw [style=simple] (28.center) to (27.center);
		\draw [style=simple] (25.center) to (20.center);
		\draw [style=simple] (18.center) to (29);
	\end{pgfonlayer}
\end{tikzpicture}
\quad\implies\quad
(\lambda x M) N \equiv M[x := N]
\end{equation}

We also recover $\eta$-reduction by the same logic. Since the variable introduced by the $\lambda$-abstraction can only occur once, the condition $x \notin \mathrm{FV}(M)$ always holds for $\lambda x M x$.

\begin{equation}
\begin{tikzpicture}
	\begin{pgfonlayer}{nodelayer}
		\node [style=none] (0) at (-2.25, -1) {};
		\node [style=none] (1) at (-1.5, -1) {$B$};
		\node [style=none] (2) at (-3.5, -0.5) {};
		\node [style=none] (3) at (-4.5, -0.75) {};
		\node [style=none] (4) at (-5.75, -0.75) {};
		\node [style=none] (5) at (-6.5, -0.75) {$\Gamma$};
		\node [style=box] (6) at (-4, -0.75) {$M$};
		\node [style=none] (7) at (-3.5, 0.5) {};
		\node [style=none] (8) at (-3, 0.5) {};
		\node [style=none] (9) at (-3, -0.5) {};
		\node [style=none] (10) at (-4.5, 1.5) {};
		\node [style=none] (11) at (-4.5, 0.5) {};
		\node [style=none] (12) at (-2.25, 1.5) {};
		\node [style=none] (13) at (0, 0) {$=$};
		\node [style=none] (14) at (-3.5, -1) {};
		\node [style=none] (15) at (-1.5, 1.5) {$A^*$};
		\node [style=none] (16) at (5.75, -0.25) {};
		\node [style=none] (17) at (6.5, -0.5) {$B$};
		\node [style=none] (18) at (4.5, 0.25) {};
		\node [style=none] (19) at (3.5, 0) {};
		\node [style=none] (20) at (2.25, 0) {};
		\node [style=none] (21) at (1.5, 0) {$\Gamma$};
		\node [style=box] (22) at (4, 0) {$M$};
		\node [style=none] (23) at (4.5, -0.25) {};
		\node [style=none] (24) at (6.75, 0.5) {$A^*$};
		\node [style=none] (25) at (5.75, 0.25) {};
	\end{pgfonlayer}
	\begin{pgfonlayer}{edgelayer}
		\draw (3.center) to (4.center);
		\draw (8.center) to (7.center);
		\draw (9.center) to (2.center);
		\draw [bend right=90, looseness=1.75] (9.center) to (8.center);
		\draw [bend left=90, looseness=1.75] (11.center) to (10.center);
		\draw [style=simple] (12.center) to (10.center);
		\draw [style=simple] (11.center) to (7.center);
		\draw [style=simple] (14.center) to (0.center);
		\draw (19.center) to (20.center);
		\draw [style=simple] (23.center) to (16.center);
		\draw [style=simple] (18.center) to (25.center);
	\end{pgfonlayer}
\end{tikzpicture}
\quad\implies\quad
\lambda x M x \equiv M
\end{equation}

\subsection{Equational rules from substitution}
In definition \ref{df:commshare} the commuting of a term through sharing was introduced. This definition presupposes the commutation rules of the ZX-calculus (principally ($\pi$) and ($c$)) shown in appendix \ref{ax:zx}. We will define some equalities on $\zeta$-terms using these.

First we look at the term $(\zeta^\alpha x M)\xi^{a \pi}$, where $a \in \{0, 1\}$. Using the basis state copy rule we show the following:

\begin{equation}
\begin{tikzpicture}
	\begin{pgfonlayer}{nodelayer}
		\node [style=none] (0) at (-3.5, 0) {};
		\node [style=none] (1) at (-2, -1.5) {};
		\node [style=none] (2) at (-1.25, -1.5) {$A$};
		\node [style=none] (3) at (-4.75, -1.25) {};
		\node [style=none] (4) at (-4.75, -1.75) {};
		\node [style=none] (5) at (-8.5, -1.75) {};
		\node [style=none] (6) at (-9.25, -1.75) {$\Gamma$};
		\node [style=box] (7) at (-4.25, -1.5) {$M$};
		\node [style=none] (8) at (-4.75, 0) {};
		\node [style=none] (9) at (-7.5, 0) {};
		\node [style=none] (10) at (-7.5, -1.25) {};
		\node [style=none] (11) at (-6.5, -0.75) {$x :_\zeta\! \underline{1}$};
		\node [style=none] (12) at (-3.5, 1.25) {};
		\node [style=none] (13) at (-3.5, 0) {};
		\node [style=none] (14) at (-3.5, 1.25) {};
		\node [style=Xp] (15) at (-6.25, 1.25) {$a\pi$};
		\node [style=Zp] (16) at (-4.25, 0) {$\alpha$};
		\node [style=none] (17) at (7, -0.25) {};
		\node [style=none] (18) at (7.5, -0.25) {$A$};
		\node [style=none] (19) at (4.5, -1) {};
		\node [style=none] (20) at (3, -1) {};
		\node [style=none] (21) at (4.5, 0.5) {};
		\node [style=Xp] (24) at (3, 0.5) {$a\pi$};
		\node [style=none] (25) at (4.5, 1) {};
		\node [style=none] (26) at (4.5, -1.5) {};
		\node [style=none] (27) at (6, -1.5) {};
		\node [style=none] (28) at (6, 1) {};
		\node [style=none] (29) at (6, -0.25) {};
		\node [style=none] (31) at (2.25, -1) {$\Gamma$};
		\node [style=none] (32) at (5.25, -0.25) {$M$};
		\node [style=none] (36) at (0, 0) {$\overset{(c)}{=}$};
	\end{pgfonlayer}
	\begin{pgfonlayer}{edgelayer}
		\draw (4.center) to (5.center);
		\draw (7) to (1.center);
		\draw (9.center) to (8.center);
		\draw (10.center) to (3.center);
		\draw [bend left=90, looseness=1.75] (10.center) to (9.center);
		\draw [bend right=90, looseness=1.75] (13.center) to (12.center);
		\draw [style=simple] (15) to (14.center);
		\draw [style=simple] (8.center) to (16);
		\draw [style=simple] (16) to (13.center);
		\draw [style=simple] (26.center)
			 to (25.center)
			 to (28.center)
			 to (27.center)
			 to cycle;
		\draw [style=simple] (24) to (21.center);
		\draw [style=simple] (20.center) to (19.center);
		\draw [style=simple] (29.center) to (17.center);
	\end{pgfonlayer}
\end{tikzpicture}
\end{equation}

By using the $(c)$ rule again we note that the term $\xi^{a \pi}$ commutes with sharing over $\zeta$ and $\emptyset$. Thus, by substitution we have the equality $(\zeta^\alpha x M)\xi^{a \pi} \equiv M[x := \xi^{a \pi}]$.

Now we look at the term $(\zeta^\alpha x M) \circ \hat{\xi}^{a \pi}$, again where $a \in \{0, 1\}$. Then we can apply the $\pi$-commutation rule to show:

\begin{equation}
\begin{tikzpicture}
	\begin{pgfonlayer}{nodelayer}
		\node [style=none] (0) at (-3.5, -0.25) {};
		\node [style=none] (1) at (-2, -1.75) {};
		\node [style=none] (2) at (-1.25, -1.75) {$B$};
		\node [style=none] (3) at (-4.75, -1.5) {};
		\node [style=none] (4) at (-4.75, -2) {};
		\node [style=none] (5) at (-8.5, -2) {};
		\node [style=none] (6) at (-9.25, -2) {$\Gamma$};
		\node [style=box] (7) at (-4.25, -1.75) {$M$};
		\node [style=none] (8) at (-4.75, -0.25) {};
		\node [style=none] (9) at (-7.5, -0.25) {};
		\node [style=none] (10) at (-7.5, -1.5) {};
		\node [style=none] (11) at (-6.5, -1) {$x :_\zeta\! A$};
		\node [style=none] (12) at (-3.5, 0.75) {};
		\node [style=none] (13) at (-3.5, -0.25) {};
		\node [style=none] (14) at (-3.5, 0.75) {};
		\node [style=Xp] (15) at (-6.25, 0.75) {$a\pi$};
		\node [style=Zp] (16) at (-4.25, -0.25) {$\alpha$};
		\node [style=none] (17) at (8.75, -0.75) {};
		\node [style=none] (18) at (9.25, -0.75) {$B$};
		\node [style=none] (19) at (6.25, -1.5) {};
		\node [style=none] (20) at (2.75, -1.5) {};
		\node [style=none] (22) at (6.25, 0.25) {};
		\node [style=Xp] (23) at (4.75, 0.25) {$a\pi$};
		\node [style=none] (25) at (6.25, 0.5) {};
		\node [style=none] (26) at (6.25, -2) {};
		\node [style=none] (27) at (7.75, -2) {};
		\node [style=none] (28) at (7.75, 0.5) {};
		\node [style=none] (29) at (7.75, -0.75) {};
		\node [style=none] (30) at (2, -1.5) {$\Gamma$};
		\node [style=none] (31) at (7, -0.75) {$M$};
		\node [style=none] (32) at (0.25, 0) {$\overset{(\pi)}{=}$};
		\node [style=none] (33) at (-7.5, 0.75) {};
		\node [style=none] (35) at (3.5, 1.5) {};
		\node [style=none] (36) at (9, 1.5) {};
		\node [style=none] (37) at (3.5, 0.25) {};
		\node [style=none] (38) at (-7.5, 1.75) {};
		\node [style=none] (39) at (-2.5, 1.75) {};
		\node [style=none] (40) at (-1.75, 1.75) {$A^*$};
		\node [style=none] (41) at (9.75, 1.5) {$A^*$};
		\node [style=Zp] (42) at (7, 1.5) {$-\alpha$};
	\end{pgfonlayer}
	\begin{pgfonlayer}{edgelayer}
		\draw (4.center) to (5.center);
		\draw (7) to (1.center);
		\draw (9.center) to (8.center);
		\draw (10.center) to (3.center);
		\draw [bend left=90, looseness=1.75] (10.center) to (9.center);
		\draw [bend right=90, looseness=1.75] (13.center) to (12.center);
		\draw [style=simple] (15) to (14.center);
		\draw [style=simple] (8.center) to (16);
		\draw [style=simple] (16) to (13.center);
		\draw [style=simple] (26.center)
			 to (25.center)
			 to (28.center)
			 to (27.center)
			 to cycle;
		\draw [style=simple] (23) to (22.center);
		\draw [style=simple] (20.center) to (19.center);
		\draw [style=simple] (29.center) to (17.center);
		\draw (33.center) to (15);
		\draw (36.center) to (35.center);
		\draw [bend left=90, looseness=1.50] (37.center) to (35.center);
		\draw [style=simple, bend right=90, looseness=2.00] (38.center) to (33.center);
		\draw [style=simple] (38.center) to (39.center);
		\draw [style=simple] (37.center) to (23);
	\end{pgfonlayer}
\end{tikzpicture}
\end{equation}

Again we note that $\hat{\xi}^{a \pi} x$ commutes with sharing over $\zeta$ and $\{x :_\zeta\! A\}$ by $\pi$-commutation (since the spiders in sharing always have phase zero). By substitution we define the equality $(\zeta^\alpha x M) \circ \hat{\xi}^{a \pi} \equiv \zeta^{\shortminus\alpha} x M[x := \hat{\xi}^{a \pi} x]$.

Lastly we will look at colour change. In the interpretation of the abstraction $\zeta^\alpha x M$, note that we can apply the colour change rule to obtain:

\begin{equation} \label{eq:sharing_comm_3}
\begin{tikzpicture}
	\begin{pgfonlayer}{nodelayer}
		\node [style=none] (0) at (-3.5, 0.5) {};
		\node [style=none] (1) at (-3.5, -1.25) {};
		\node [style=none] (2) at (-3, 0.5) {$A^*$};
		\node [style=none] (3) at (-3, -1.25) {$B$};
		\node [style=none] (4) at (-5.5, -1) {};
		\node [style=none] (5) at (-5.5, -1.5) {};
		\node [style=none] (6) at (-7.5, -1.5) {};
		\node [style=none] (7) at (-8.25, -1.5) {$\Gamma$};
		\node [style=box] (8) at (-5, -1.25) {$M$};
		\node [style=none] (9) at (-5.5, 0.5) {};
		\node [style=none] (10) at (-6, 0.5) {};
		\node [style=none] (11) at (-6, -1) {};
		\node [style=none] (12) at (6.75, -1.25) {};
		\node [style=none] (13) at (7.25, -1.25) {$B$};
		\node [style=none] (14) at (5, -1) {};
		\node [style=none] (15) at (5, -1.5) {};
		\node [style=none] (16) at (2.5, -1.5) {};
		\node [style=none] (17) at (1.75, -1.5) {$\Gamma$};
		\node [style=box] (18) at (5.5, -1.25) {$M$};
		\node [style=none] (19) at (3.5, -0.25) {};
		\node [style=none] (20) at (3.5, -1) {};
		\node [style=none] (21) at (0, 0) {$\overset{(h)}{=}$};
		\node [style=Zp] (22) at (-5, 0.5) {$\alpha$};
		\node [style=H] (23) at (4.25, -1) {};
		\node [style=Xp] (24) at (4.25, -0.25) {$\alpha$};
		\node [style=H] (25) at (4.25, 0.5) {};
		\node [style=none] (26) at (5, -0.25) {};
		\node [style=none] (27) at (5, 0.5) {};
		\node [style=none] (28) at (3.5, 0.5) {};
		\node [style=none] (29) at (3.5, 1.25) {};
		\node [style=none] (30) at (5, 1.25) {};
		\node [style=none] (31) at (6.75, 1.25) {};
		\node [style=none] (32) at (7.25, 1.25) {$A^*$};
	\end{pgfonlayer}
	\begin{pgfonlayer}{edgelayer}
		\draw (5.center) to (6.center);
		\draw (8) to (1.center);
		\draw (10.center) to (9.center);
		\draw (11.center) to (4.center);
		\draw [bend left=90, looseness=1.75] (11.center) to (10.center);
		\draw (15.center) to (16.center);
		\draw (18) to (12.center);
		\draw [bend left=90, looseness=1.75] (20.center) to (19.center);
		\draw (9.center) to (22);
		\draw (22) to (0.center);
		\draw (19.center) to (24);
		\draw (20.center) to (23);
		\draw (23) to (14.center);
		\draw (24) to (26.center);
		\draw [bend right=90, looseness=1.75] (26.center) to (27.center);
		\draw (28.center) to (25);
		\draw (25) to (27.center);
		\draw [in=-180, out=180, looseness=1.75] (28.center) to (29.center);
		\draw (29.center) to (30.center);
		\draw (31.center) to (30.center);
	\end{pgfonlayer}
\end{tikzpicture}
\end{equation}

The term $H x$ commutes with sharing over $\zeta$ and $\{x :_\xi\! A\}$, switching the basis which $x$ is introduced in by Hadamard pushing through the sharing spiders. Looking again at equation \eqref{eq:sharing_comm_3} we see that this is expected since the abstraction has changed its basis. By substitution we obtain the term $\xi^\alpha x M[x := H x]$, which we then compose with $H$ to reconstruct the original diagram. With this we get the colour change rule in $\zeta$ as $\zeta^\alpha x M \equiv (\xi^\alpha x M[x := H x]) \circ H$.

\subsection{The full theory}

The remaining equational rules relate to spider fusion together with more trivial rules on simple terms. Combining these with the rules introduced in the previous sections we present the full equational theory of the $\zeta$-calculus in figure \ref{fig:eqth}.

\begin{figure}[h]
\[
\begin{aligned}[t]
\beta x M &\equiv \lambda x M \qquad (\omega_x(M) = 1) \\
(\lambda x M) N &\equiv M[x := N] \\
\lambda x M x &\equiv M \\
(\zeta^\alpha x M) \xi^{a \pi} &\equiv M[x := \xi^{a \pi}] \\
(\zeta^\alpha x M) \circ \hat{\xi}^{a \pi} &\equiv \zeta^{-\alpha} x M[x := \hat{\xi}^{a \pi} x] \\
\zeta^\alpha x M &\equiv (\xi^{\alpha} x M[x := H x]) \circ H\\
\end{aligned}
\qquad\qquad
\begin{aligned}[t]
\beta^\alpha x M &\equiv \beta^\alpha x N \qquad (M \equiv N) \\
(\beta^\alpha x M) \beta^\theta &\equiv (\beta x M) \beta^{\theta + \alpha} \\
(\beta^\alpha x M) \circ \hat{\beta}^\theta &\equiv \beta^{\alpha + \theta} x M \\
H \zeta^\alpha_n &\equiv \xi^\alpha_n \\
H (H M) &\equiv M \\
\langle M, * \rangle &\equiv M \\
\langle *, M \rangle &\equiv M
\end{aligned}
\]
\caption{Rules of the equational theory.}
\label{fig:eqth}
\end{figure}

The equational theory presented here is \emph{not} complete. To make it so a more comprehensive theory of sharing is needed, and possibly more. This problem will be highlighted in section \ref{sec:examples} where we rely on translations between ZX and $\zeta$ instead. Nevertheless, the equational theory is sound shown by the following theorem.

\begin{theorem}[Soundness]
$\zeta \vdash M \equiv N \implies ZX \vdash \llbracket \Gamma \vdash M : A \rrbracket = \llbracket \Gamma \vdash N : A \rrbracket$
\end{theorem}
\begin{proof}
The proof is by showing equality of the interpretation of each of the rules in the equational theory of the ZX-calculus.
See appendix \ref{appendix:soundness} for the full proof.

\end{proof}

\section{Examples} \label{sec:examples}

To highlight the main features of the $\zeta$-calculus we present some illustrative examples in this section. The focus is on common constructs of the ZX-calculus as well as higher-order functions and sharing. To give some intuition on how $\zeta$-terms are represented in ZX, we introduce the relation $M \cong D$ on closed $\zeta$-terms $M$ and ZX-diagrams with boxes $D$. The intention is that this shows the interpretation $\llbracket \vdash M : A \rrbracket$ as a diagram, possibly with arbitrary terms represented by boxes, with function types "externalised". That is, reversing the dual types into inputs. A simple example of this would be:
\begin{equation}
\llbracket\; \vdash \zeta^\alpha x M \;:\; \underline{1} \to \underline{1} \;\rrbracket \quad=\quad \begin{tikzpicture}
	\begin{pgfonlayer}{nodelayer}
		\node [style=box] (0) at (0, -0.75) {$M$};
		\node [style=none] (1) at (1.25, -0.75) {};
		\node [style=Zp] (2) at (0, 0.75) {$\alpha$};
		\node [style=none] (3) at (1.25, 0.75) {};
		\node [style=none] (4) at (-1, 0.75) {};
		\node [style=none] (5) at (-1, -0.75) {};
		\node [style=none] (6) at (2, 0.75) {$\underline{1}^*$};
		\node [style=none] (7) at (1.75, -0.75) {$\underline{1}$};
		\node [style=none] (8) at (3.5, 0) {$\rightsquigarrow$};
		\node [style=box] (9) at (7, 0) {$M$};
		\node [style=none] (10) at (8.25, 0) {};
		\node [style=Zp] (13) at (5.75, 0) {$\alpha$};
		\node [style=none] (14) at (4.75, 0) {};
	\end{pgfonlayer}
	\begin{pgfonlayer}{edgelayer}
		\draw [style=simple] (2) to (3.center);
		\draw [style=simple] (2) to (4.center);
		\draw [style=simple, bend right=90, looseness=1.75] (4.center) to (5.center);
		\draw [style=simple] (5.center) to (0);
		\draw [style=simple] (0) to (1.center);
		\draw [style=simple] (9) to (10.center);
		\draw [style=simple] (13) to (14.center);
		\draw [style=simple] (13) to (9);
	\end{pgfonlayer}
\end{tikzpicture}
\end{equation}
For which we would then write $\zeta^\alpha x M \;\cong\; \begin{tikzpicture}
	\begin{pgfonlayer}{nodelayer}
		\node [style=box] (9) at (0.5, 0) {$M$};
		\node [style=none] (10) at (1.75, 0) {};
		\node [style=Zp] (13) at (-0.75, 0) {$\alpha$};
		\node [style=none] (14) at (-1.75, 0) {};
	\end{pgfonlayer}
	\begin{pgfonlayer}{edgelayer}
		\draw [style=simple] (9) to (10.center);
		\draw [style=simple] (13) to (14.center);
		\draw [style=simple] (13) to (9);
	\end{pgfonlayer}
\end{tikzpicture}$. This provides a clear way to interpret for instance the action of a higher order function in the $\zeta$-calculus as a ZX-diagram.

\subsection{Phase gadgets, linking functions, and multi-qubit unitaries}

Phase gadgets are ZX-diagrams of the form \eqref{eq:phase_gadget} implementing the action of a unitary operator $U_f$ on a string of input wires $\ket{\Vec{x}}:=\ket{x_1 \dots x_n}$ as $U_f \ket{\Vec{x}} = e^{i f(\Vec{x})} \ket{\Vec{x}}$, where $f(\Vec{x}) = \alpha\;(x_1 \oplus \dots \oplus x_n)$ for some phase $\alpha$ \cite{vandeWetering2020ZX-calculusScientist}.
%
\begin{equation} \label{eq:phase_gadget}
\begin{tikzpicture}
	\begin{pgfonlayer}{nodelayer}
		\node [style=none] (0) at (0, 0) {};
		\node [style=none] (1) at (4, 0) {};
		\node [style=Z] (2) at (1.5, 0) {};
		\node [style=none] (3) at (0, -2.5) {};
		\node [style=none] (4) at (4, -2.5) {};
		\node [style=Z] (5) at (1.5, -2.5) {};
		\node [style=none] (6) at (0, -1) {};
		\node [style=none] (7) at (4, -1) {};
		\node [style=Z] (8) at (1.5, -1) {};
		\node [style=none] (9) at (3.25, -1.5) {$\vdots$};
		\node [style=X] (10) at (3, 1) {};
		\node [style=Zp] (11) at (3, 2) {$\alpha$};
	\end{pgfonlayer}
	\begin{pgfonlayer}{edgelayer}
		\draw [style=simple] (0.center) to (2);
		\draw [style=simple] (2) to (1.center);
		\draw [style=simple] (3.center) to (5);
		\draw [style=simple] (5) to (4.center);
		\draw [style=simple] (6.center) to (8);
		\draw [style=simple] (8) to (7.center);
		\draw [style=simple] (2) to (10);
		\draw [style=simple] (8) to (10);
		\draw [style=simple] (5) to (10);
		\draw [style=simple] (10) to (11);
	\end{pgfonlayer}
\end{tikzpicture}
\end{equation}
We can implement phase gadgets that act on the inputs of a basis abstraction by supplying shared arguments to a gadget function $G^\alpha : \underline{n} \to \top$,
\begin{equation}
G^\alpha \;:\equiv\; \lambda x \; \xi_k \langle \zeta^\alpha, x \rangle \quad\cong\quad \begin{tikzpicture}
	\begin{pgfonlayer}{nodelayer}
		\node [style=none] (3) at (0, -1) {};
		\node [style=none] (4) at (0, 0) {};
		\node [style=Zp] (5) at (0, 1) {$\alpha$};
		\node [style=X] (6) at (1.25, 0) {};
		\node [style=none] (7) at (-0.75, -1) {};
		\node [style=none] (8) at (-0.75, 0) {};
		\node [style=none] (9) at (-0.25, -0.25) {$\vdots$};
		\node [style=none] (10) at (-1, -1) {};
		\node [style=none] (11) at (-1, 0) {};
		\node [style=none] (12) at (-1.5, -0.5) {$n$};
	\end{pgfonlayer}
	\begin{pgfonlayer}{edgelayer}
		\draw [in=-150, out=0, looseness=1.25] (3.center) to (6);
		\draw (4.center) to (6);
		\draw [in=150, out=0, looseness=1.25] (5) to (6);
		\draw (3.center) to (7.center);
		\draw (4.center) to (8.center);
		\draw [style=brace edge] (10.center) to (11.center);
	\end{pgfonlayer}
\end{tikzpicture} \qquad \mathrm{where}\; k = - n - 1 .
\end{equation}
We can then use this gadget function together with any basis abstraction to add a phase to the inputs depending on their parity,
\begin{equation}
\zeta x \zeta y \; \langle G^\alpha \langle x, y \rangle, M \rangle \quad\cong\quad \begin{tikzpicture}
	\begin{pgfonlayer}{nodelayer}
		\node [style=none] (1) at (1.5, 0.5) {};
		\node [style=none] (2) at (1.5, -0.5) {};
		\node [style=Z] (3) at (0.75, 0.5) {};
		\node [style=Z] (4) at (0.75, -0.5) {};
		\node [style=X] (5) at (1.5, 1.5) {};
		\node [style=Zp] (6) at (2.25, 1.5) {$\alpha$};
		\node [style=none] (7) at (3.5, 0) {};
		\node [style=none] (8) at (0, 0.5) {};
		\node [style=none] (9) at (0, -0.5) {};
		\node [style=none] (10) at (1.5, 0.75) {};
		\node [style=none] (11) at (2.5, 0.75) {};
		\node [style=none] (12) at (2.5, -0.75) {};
		\node [style=none] (13) at (1.5, -0.75) {};
		\node [style=none] (14) at (2.5, 0) {};
		\node [style=none] (15) at (2, 0) {$M$};
	\end{pgfonlayer}
	\begin{pgfonlayer}{edgelayer}
		\draw [style=simple] (8.center) to (3);
		\draw [style=simple] (9.center) to (4);
		\draw [style=simple] (3) to (1.center);
		\draw [style=simple] (4) to (2.center);
		\draw [style=simple] (3) to (5);
		\draw [style=simple] (4) to (5);
		\draw [style=simple] (5) to (6);
		\draw [style=simple] (10.center)
			 to (11.center)
			 to (12.center)
			 to (13.center)
			 to cycle;
		\draw [style=simple] (14.center) to (7.center);
	\end{pgfonlayer}
\end{tikzpicture} \qquad.
\end{equation}
We can use functions on this form in general to link variables introduced in a basis. The function that links two variables through identity becomes $L :\equiv \zeta_{\shortminus 2} \;\cong\; \begin{tikzpicture}
	\begin{pgfonlayer}{nodelayer}
		\node [style=Z] (0) at (0, 0) {};
		\node [style=none] (1) at (-1, 0.5) {};
		\node [style=none] (2) at (-1, -0.5) {};
	\end{pgfonlayer}
	\begin{pgfonlayer}{edgelayer}
		\draw [style=simple, in=165, out=0, looseness=1.25] (1.center) to (0);
		\draw [style=simple, in=0, out=-165, looseness=1.25] (0) to (2.center);
	\end{pgfonlayer}
\end{tikzpicture}$, while linking through Hadamard becomes $L_H :\equiv \zeta^{\frac{\pi}{2}}_{\shortminus 2} \circ \hat{\xi}^{\frac{\pi}{2}} \;\cong\; \begin{tikzpicture}
	\begin{pgfonlayer}{nodelayer}
		\node [style=Zp] (0) at (0.25, 0) {$\frac{\pi}{2}$};
		\node [style=Xp] (1) at (-0.75, 0.5) {$\frac{\pi}{2}$};
		\node [style=Xp] (2) at (-0.75, -0.5) {$\frac{\pi}{2}$};
		\node [style=none] (3) at (-1.5, 0.5) {};
		\node [style=none] (4) at (-1.5, -0.5) {};
	\end{pgfonlayer}
	\begin{pgfonlayer}{edgelayer}
		\draw [style=simple] (3.center) to (1);
		\draw [style=simple] (4.center) to (2);
		\draw [style=simple, in=165, out=0, looseness=1.25] (1) to (0);
		\draw [style=simple, in=0, out=-165, looseness=1.25] (0) to (2);
	\end{pgfonlayer}
\end{tikzpicture}$. Using \textit{linking functions} like these makes for a concise way to construct multi-qubit unitaries, for example $CNOT \equiv \zeta c \xi t \langle L \langle c, t \rangle, c , t \rangle$ and $C_z \equiv \zeta c \zeta t \langle L_H \langle c, t \rangle, c, t \rangle$.

\subsection{Higher-order functions via sharing}

To elaborate further on the features of $\zeta$ illustrated in section \ref{subsec:hos} we will look at a function using higher-order sharing. We define the \textit{Pauli switching} function $switch_\beta : (A \to A) \to A \to A$ in \eqref{eq:switch}.
\begin{equation} \label{eq:switch}
\llbracket\; \vdash \beta f \; f \circ H \circ f \;:\; (A \to A) \to A \to A \;\rrbracket \quad=\quad \begin{tikzpicture}
	\begin{pgfonlayer}{nodelayer}
		\node [style=T] (0) at (-0.25, 1.5) {};
		\node [style=T] (1) at (-0.25, 0.5) {};
		\node [style=none] (2) at (1.75, 1.5) {};
		\node [style=none] (3) at (1.75, 0.5) {};
		\node [style=none] (4) at (1.75, -0.5) {};
		\node [style=none] (5) at (1.75, -1.5) {};
		\node [style=none] (9) at (-0.25, -0.5) {};
		\node [style=none] (10) at (-0.25, -1.5) {};
		\node [style=H] (11) at (-0.25, 1) {};
		\node [style=none] (12) at (2.25, 1.5) {$A$};
		\node [style=none] (13) at (2.5, 0.5) {$A^*$};
		\node [style=none] (14) at (2.5, -0.5) {$A^*$};
		\node [style=none] (15) at (2.25, -1.5) {$A$};
		\node [style=none] (16) at (3.25, 1.75) {};
		\node [style=none] (17) at (3.25, 0.25) {};
		\node [style=none] (18) at (3.25, -0.25) {};
		\node [style=none] (19) at (3.25, -1.75) {};
		\node [style=none] (20) at (5, 1) {$(A \to A)^*$};
		\node [style=none] (21) at (4.75, -1) {$A \to A$};
		\node [style=none] (22) at (6.75, 1.5) {};
		\node [style=none] (23) at (6.75, -1.5) {};
		\node [style=none] (24) at (10, 0) {$(A \to A) \to A \to A$};
	\end{pgfonlayer}
	\begin{pgfonlayer}{edgelayer}
		\draw [style=simple] (0) to (11);
		\draw [style=simple] (11) to (1);
		\draw [style=simple] (0) to (2.center);
		\draw [style=simple] (1) to (3.center);
		\draw [style=simple, in=180, out=-180, looseness=2.00] (1) to (10.center);
		\draw [style=simple, in=180, out=-180, looseness=2.00] (0) to (9.center);
		\draw [style=simple] (9.center) to (4.center);
		\draw [style=simple] (10.center) to (5.center);
		\draw [style=brace edge] (16.center) to (17.center);
		\draw [style=brace edge] (18.center) to (19.center);
		\draw [style=brace edge] (22.center) to (23.center);
	\end{pgfonlayer}
\end{tikzpicture}
\end{equation}
Applying $switch_\beta$ to some term $M : A \to A$ we can view it in a more digestible fashion:
\begin{equation}
switch_\beta \; M \quad\cong\quad\begin{tikzpicture}
	\begin{pgfonlayer}{nodelayer}
		\node [style=box] (0) at (0, 0) {$M$};
		\node [style=T] (1) at (1.25, 0) {};
		\node [style=T] (2) at (-1.25, 0) {};
		\node [style=H] (3) at (0, 1.25) {};
		\node [style=none] (4) at (-2, 0) {};
		\node [style=none] (5) at (2, 0) {};
	\end{pgfonlayer}
	\begin{pgfonlayer}{edgelayer}
		\draw [style=simple] (4.center) to (2);
		\draw [style=simple] (1) to (5.center);
		\draw [style=simple] (2) to (0);
		\draw [style=simple] (0) to (1);
		\draw [style=simple, in=-180, out=90, looseness=1.25] (2) to (3);
		\draw [style=simple, in=90, out=0, looseness=1.25] (3) to (1);
	\end{pgfonlayer}
\end{tikzpicture}
\end{equation}
We call it the Pauli switching function because of its behaviour when applied to the Pauli gates ($\sigma_x \equiv \hat{\xi}^\pi$, $\sigma_z \equiv \hat{\zeta}^\pi$, and $\sigma_y \equiv \hat{\xi}^\pi \circ \hat{\zeta}^\pi$). When applied to the identity function it "switches on" the Pauli gate of the basis it is shared through.
\begin{equation}
switch_\beta \; (\lambda x x) \quad\cong\quad \begin{tikzpicture}
	\begin{pgfonlayer}{nodelayer}
		\node [style=T] (1) at (1, 0) {};
		\node [style=T] (2) at (-0.5, 0) {};
		\node [style=H] (3) at (0.25, 0.75) {};
		\node [style=none] (4) at (-1.5, 0) {};
		\node [style=none] (5) at (2, 0) {};
		\node [style=none] (6) at (3, 0.25) {$\overset{(f)}{=}$};
		\node [style=T] (7) at (5, 0) {};
		\node [style=none] (8) at (4, 0) {};
		\node [style=none] (9) at (6, 0) {};
		\node [style=H] (10) at (5, 0.75) {};
		\node [style=none] (11) at (7, 0) {$=$};
		\node [style=Tp] (12) at (9, 0) {$\pi$};
		\node [style=none] (13) at (8, 0) {};
		\node [style=none] (14) at (10, 0) {};
	\end{pgfonlayer}
	\begin{pgfonlayer}{edgelayer}
		\draw [style=simple] (4.center) to (2);
		\draw [style=simple] (1) to (5.center);
		\draw [style=simple, in=-180, out=90, looseness=1.25] (2) to (3);
		\draw [style=simple, in=90, out=0, looseness=1.25] (3) to (1);
		\draw [style=simple] (2) to (1);
		\draw [style=simple, in=-165, out=150, looseness=1.50] (7) to (10);
		\draw [style=simple] (8.center) to (7);
		\draw [style=simple] (7) to (9.center);
		\draw [style=simple, in=30, out=-15, looseness=1.50] (10) to (7);
		\draw [style=simple] (13.center) to (12);
		\draw [style=simple] (12) to (14.center);
	\end{pgfonlayer}
\end{tikzpicture} \quad\cong\quad \hat{\beta}^\pi
\end{equation}
Then, applying $switch_\beta$ to the Pauli gate in the basis $\beta$ switches it off.
\begin{equation}
switch_\beta \; \hat{\beta}^\pi \quad\cong\quad \begin{tikzpicture}
	\begin{pgfonlayer}{nodelayer}
		\node [style=T] (1) at (1.25, 0) {};
		\node [style=T] (2) at (-1.25, 0) {};
		\node [style=H] (3) at (0, 1.25) {};
		\node [style=none] (4) at (-2, 0) {};
		\node [style=none] (5) at (2, 0) {};
		\node [style=none] (6) at (3, 0.25) {$\overset{(f)}{=}$};
		\node [style=none] (8) at (4, 0) {};
		\node [style=none] (9) at (6, 0) {};
		\node [style=H] (10) at (5, 1) {};
		\node [style=none] (11) at (7, 0) {$=$};
		\node [style=none] (13) at (8, 0) {};
		\node [style=none] (14) at (10, 0) {};
		\node [style=Tp] (15) at (0, 0) {$\pi$};
		\node [style=Tp] (16) at (5, 0) {$\pi$};
		\node [style=T] (17) at (9, 0) {};
		\node [style=none] (18) at (11, 0.25) {$\overset{(id)}{=}$};
		\node [style=none] (19) at (12, 0) {};
		\node [style=none] (20) at (14, 0) {};
	\end{pgfonlayer}
	\begin{pgfonlayer}{edgelayer}
		\draw [style=simple] (4.center) to (2);
		\draw [style=simple] (1) to (5.center);
		\draw [style=simple, in=-180, out=90, looseness=1.25] (2) to (3);
		\draw [style=simple, in=90, out=0, looseness=1.25] (3) to (1);
		\draw [style=simple, in=180, out=0] (2) to (15);
		\draw [style=simple] (15) to (1);
		\draw [style=simple] (8.center) to (16);
		\draw [style=simple] (16) to (9.center);
		\draw [style=simple, in=-15, out=15, looseness=1.50] (16) to (10);
		\draw [style=simple, in=165, out=-165, looseness=1.50] (10) to (16);
		\draw [style=simple] (13.center) to (17);
		\draw [style=simple] (17) to (14.center);
		\draw [style=simple] (19.center) to (20.center);
	\end{pgfonlayer}
\end{tikzpicture} \quad\cong\quad \lambda x x
\end{equation}
We can see that this behaviour is expected more generally since the function is self-inverse \eqref{eq:switch_inverse}.
\begin{equation} \label{eq:switch_inverse}
switch_\beta^{\circ 2} \; M \quad\cong\quad \begin{tikzpicture}
	\begin{pgfonlayer}{nodelayer}
		\node [style=box] (0) at (-7.25, 0) {$M$};
		\node [style=T] (1) at (-6, 0) {};
		\node [style=T] (2) at (-8.5, 0) {};
		\node [style=H] (3) at (-7.25, 1.25) {};
		\node [style=T] (4) at (-9.25, 0) {};
		\node [style=T] (5) at (-5.25, 0) {};
		\node [style=H] (6) at (-7.25, -1.25) {};
		\node [style=none] (7) at (-4.25, 0) {};
		\node [style=none] (8) at (-10.25, 0) {};
		\node [style=none] (9) at (-3.25, 0.25) {$\overset{(f)}{=}$};
		\node [style=box] (10) at (0, 0) {$M$};
		\node [style=T] (11) at (1.25, 0) {};
		\node [style=T] (12) at (-1.25, 0) {};
		\node [style=H] (13) at (0, 1.25) {};
		\node [style=H] (16) at (0, -1.25) {};
		\node [style=none] (17) at (2.25, 0) {};
		\node [style=none] (18) at (-2.25, 0) {};
		\node [style=none] (19) at (3.25, 0) {$=$};
		\node [style=box] (20) at (6.5, 0) {$M$};
		\node [style=T] (21) at (7.75, 0) {};
		\node [style=T] (22) at (5.25, 0) {};
		\node [style=none] (27) at (8.75, 0) {};
		\node [style=none] (28) at (4.25, 0) {};
		\node [style=none] (29) at (9.75, 0.25) {$\overset{(id)}{=}$};
		\node [style=box] (30) at (11.75, 0) {$M$};
		\node [style=none] (35) at (12.75, 0) {};
		\node [style=none] (36) at (10.75, 0) {};
	\end{pgfonlayer}
	\begin{pgfonlayer}{edgelayer}
		\draw [style=simple] (2) to (0);
		\draw [style=simple] (0) to (1);
		\draw [style=simple, in=-180, out=90, looseness=1.25] (2) to (3);
		\draw [style=simple, in=90, out=0, looseness=1.25] (3) to (1);
		\draw [in=180, out=-90, looseness=1.25] (4) to (6);
		\draw [in=-90, out=0, looseness=1.25] (6) to (5);
		\draw [style=simple] (7.center) to (5);
		\draw [style=simple] (5) to (1);
		\draw [style=simple] (8.center) to (4);
		\draw [style=simple] (4) to (2);
		\draw [style=simple] (12) to (10);
		\draw [style=simple] (10) to (11);
		\draw [style=simple, in=-180, out=90, looseness=1.25] (12) to (13);
		\draw [style=simple, in=90, out=0, looseness=1.25] (13) to (11);
		\draw [style=simple] (22) to (20);
		\draw [style=simple] (20) to (21);
		\draw [style=simple] (36.center) to (30);
		\draw [style=simple] (30) to (35.center);
		\draw [style=simple] (18.center) to (12);
		\draw [style=simple] (11) to (17.center);
		\draw [style=simple, in=-180, out=-90, looseness=1.25] (12) to (16);
		\draw [style=simple, in=-90, out=0, looseness=1.25] (16) to (11);
		\draw [style=simple] (28.center) to (22);
		\draw [style=simple] (21) to (27.center);
	\end{pgfonlayer}
\end{tikzpicture}
\end{equation}
This function also illustrates the difference that sharing in different bases makes. Applying it to a Pauli gate of a different basis does nothing.
\begin{equation}
\begin{tikzpicture}
	\begin{pgfonlayer}{nodelayer}
		\node [style=H] (3) at (0, 1) {};
		\node [style=none] (4) at (-2, 0) {};
		\node [style=none] (5) at (2, 0) {};
		\node [style=Z] (6) at (-1, 0) {};
		\node [style=Z] (7) at (1, 0) {};
		\node [style=Xp] (8) at (0, 0) {$\pi$};
		\node [style=none] (9) at (3, 0.25) {$\overset{(\pi)}{=}$};
		\node [style=none] (10) at (4, 0) {};
		\node [style=Z] (11) at (5, 0) {};
		\node [style=H] (12) at (6, 1) {};
		\node [style=Xp] (13) at (7, 1) {$\pi$};
		\node [style=Z] (14) at (8, 0) {};
		\node [style=Xp] (15) at (9, 0) {$\pi$};
		\node [style=none] (16) at (10, 0) {};
		\node [style=none] (17) at (11, 0.25) {$\overset{(f)}{=}$};
		\node [style=none] (18) at (12, 0) {};
		\node [style=Z] (19) at (13.25, 0) {};
		\node [style=H] (20) at (12.75, 0.75) {};
		\node [style=Xp] (21) at (13.75, 0.75) {$\pi$};
		\node [style=Xp] (23) at (14.75, 0) {$\pi$};
		\node [style=none] (24) at (15.75, 0) {};
		\node [style=none] (25) at (-5.5, -2.5) {$\overset{(h)}{=}$};
		\node [style=none] (26) at (-4.5, -2.75) {};
		\node [style=H] (29) at (-2.75, -1.75) {};
		\node [style=Xp] (31) at (-3.75, -1.75) {$\pi$};
		\node [style=Xp] (32) at (-2.25, -2.75) {$\pi$};
		\node [style=none] (33) at (-1.25, -2.75) {};
		\node [style=none] (58) at (10.75, -2.75) {};
		\node [style=Xp] (61) at (11.75, -2.75) {$\pi$};
		\node [style=none] (62) at (12.75, -2.75) {};
		\node [style=none] (63) at (9.75, -2.5) {$\overset{(id)}{=}$};
		\node [style=none] (64) at (-3, 0) {$\cong$};
		\node [style=none] (65) at (-5.75, 0) {$switch_\zeta \; \hat{\xi}^\pi$};
		\node [style=none] (66) at (14, -2.75) {$\cong$};
		\node [style=none] (67) at (15.5, -2.75) {$\hat{\xi}^\pi$};
		\node [style=Zp] (69) at (-3.75, -1.75) {$\pi$};
		\node [style=none] (70) at (-0.25, -2.5) {$\overset{(f)}{=}$};
		\node [style=none] (71) at (0.75, -2.75) {};
		\node [style=Zp] (72) at (1.75, -2.75) {$\pi$};
		\node [style=Xp] (73) at (2.75, -2.75) {$\pi$};
		\node [style=none] (74) at (3.75, -2.75) {};
		\node [style=H] (75) at (1.75, -2) {};
		\node [style=none] (76) at (4.75, -2.75) {$=$};
		\node [style=none] (77) at (5.75, -2.75) {};
		\node [style=Z] (78) at (6.75, -2.75) {};
		\node [style=Xp] (79) at (7.75, -2.75) {$\pi$};
		\node [style=none] (80) at (8.75, -2.75) {};
		\node [style=Z] (81) at (-3.25, -2.75) {};
	\end{pgfonlayer}
	\begin{pgfonlayer}{edgelayer}
		\draw [style=simple] (4.center) to (6);
		\draw [style=simple] (6) to (8);
		\draw [style=simple] (8) to (7);
		\draw [style=simple] (7) to (5.center);
		\draw [style=simple, in=-180, out=90, looseness=1.25] (6) to (3);
		\draw [style=simple, in=90, out=0, looseness=1.25] (3) to (7);
		\draw [style=simple] (10.center) to (11);
		\draw [style=simple] (11) to (14);
		\draw [style=simple] (14) to (15);
		\draw [style=simple] (15) to (16.center);
		\draw [style=simple, in=-180, out=90, looseness=1.25] (11) to (12);
		\draw [style=simple] (12) to (13);
		\draw [style=simple, in=90, out=0, looseness=1.25] (13) to (14);
		\draw [style=simple] (18.center) to (19);
		\draw [style=simple] (23) to (24.center);
		\draw [style=simple, in=-90, out=165, looseness=1.25] (19) to (20);
		\draw [style=simple, bend left=90, looseness=1.75] (20) to (21);
		\draw [style=simple] (32) to (33.center);
		\draw [style=simple, bend left=90, looseness=1.75] (31) to (29);
		\draw [style=simple] (61) to (62.center);
		\draw [style=simple] (58.center) to (61);
		\draw [style=simple] (71.center) to (72);
		\draw [style=simple] (72) to (73);
		\draw [style=simple] (73) to (74.center);
		\draw [style=simple, bend left=90, looseness=1.75] (72) to (75);
		\draw [style=simple, bend left=90, looseness=1.75] (75) to (72);
		\draw [style=simple] (77.center) to (78);
		\draw [style=simple] (78) to (79);
		\draw [style=simple] (79) to (80.center);
		\draw [style=simple, in=-90, out=15, looseness=1.25] (19) to (21);
		\draw [style=simple] (19) to (23);
		\draw [style=simple] (26.center) to (81);
		\draw [style=simple] (81) to (32);
		\draw [style=simple, in=-90, out=30, looseness=1.25] (81) to (29);
		\draw [style=simple, in=-90, out=150, looseness=1.25] (81) to (69);
	\end{pgfonlayer}
\end{tikzpicture}
\end{equation}
\begin{equation}
\begin{tikzpicture}
	\begin{pgfonlayer}{nodelayer}
		\node [style=H] (3) at (0, 1) {};
		\node [style=none] (4) at (-2, 0) {};
		\node [style=none] (5) at (2, 0) {};
		\node [style=Z] (6) at (-1.25, 0) {};
		\node [style=Z] (7) at (1.25, 0) {};
		\node [style=Xp] (8) at (0.5, 0) {$\pi$};
		\node [style=none] (9) at (3, 0.25) {$\overset{(\pi, f)}{=}$};
		\node [style=none] (10) at (4, 0) {};
		\node [style=H] (12) at (6, 1) {};
		\node [style=Xp] (13) at (7, 1) {$\pi$};
		\node [style=Z] (14) at (8, 0) {};
		\node [style=Xp] (15) at (9, 0) {$\pi$};
		\node [style=none] (16) at (10, 0) {};
		\node [style=none] (17) at (11, 0.25) {$\overset{(f)}{=}$};
		\node [style=none] (18) at (12, 0) {};
		\node [style=H] (20) at (12.75, 1) {};
		\node [style=Xp] (21) at (13.75, 1) {$\pi$};
		\node [style=Xp] (23) at (14.75, 0) {$\pi$};
		\node [style=none] (24) at (15.75, 0) {};
		\node [style=none] (25) at (-10.75, -2.25) {$\overset{(h)}{=}$};
		\node [style=none] (26) at (-9.75, -2.5) {};
		\node [style=H] (27) at (-8.75, -2.5) {};
		\node [style=H] (28) at (-6.75, -2.5) {};
		\node [style=H] (29) at (-7.25, -1.5) {};
		\node [style=Xp] (31) at (-8.25, -1.5) {$\pi$};
		\node [style=Xp] (32) at (-5.75, -2.5) {$\pi$};
		\node [style=none] (33) at (-4.75, -2.5) {};
		\node [style=none] (34) at (-2.75, -2.5) {};
		\node [style=H] (35) at (-1.75, -2.5) {};
		\node [style=H] (36) at (0.25, -2.5) {};
		\node [style=H] (37) at (-0.75, -1.5) {};
		\node [style=Xp] (40) at (1.25, -2.5) {$\pi$};
		\node [style=none] (41) at (2.25, -2.5) {};
		\node [style=none] (42) at (-3.75, -2.25) {$\overset{(f)}{=}$};
		\node [style=none] (44) at (4.25, -2.5) {};
		\node [style=H] (45) at (5.25, -2.5) {};
		\node [style=H] (46) at (7.25, -2.5) {};
		\node [style=Xp] (48) at (8.25, -2.5) {$\pi$};
		\node [style=none] (49) at (9.25, -2.5) {};
		\node [style=none] (50) at (3.25, -2.5) {$=$};
		\node [style=none] (57) at (10.25, -2.25) {$\overset{(h)}{=}$};
		\node [style=Zp] (64) at (-0.5, 0) {$\pi$};
		\node [style=Zp] (65) at (5, 0) {$\pi$};
		\node [style=Zp] (66) at (13.25, 0) {$\pi$};
		\node [style=Xp] (67) at (-7.75, -2.5) {$\pi$};
		\node [style=X] (68) at (-0.75, -2.5) {};
		\node [style=Xp] (69) at (6.25, -2.5) {$\pi$};
		\node [style=none] (71) at (11.5, -2.5) {};
		\node [style=none] (72) at (14.5, -2.5) {};
		\node [style=Xp] (75) at (13.5, -2.5) {$\pi$};
		\node [style=Zp] (76) at (12.5, -2.5) {$\pi$};
		\node [style=none] (77) at (-3, 0) {$\cong$};
		\node [style=none] (78) at (-6.75, 0) {$switch_\zeta \; (\hat{\xi}^\pi \circ \hat{\zeta}^\pi)$};
		\node [style=none] (79) at (15.5, -2.5) {$\cong$};
		\node [style=none] (80) at (17.75, -2.5) {$\hat{\xi}^\pi \circ \hat{\zeta}^\pi$};
	\end{pgfonlayer}
	\begin{pgfonlayer}{edgelayer}
		\draw [style=simple] (4.center) to (6);
		\draw [style=simple] (6) to (8);
		\draw [style=simple] (8) to (7);
		\draw [style=simple] (7) to (5.center);
		\draw [style=simple, in=-180, out=90, looseness=1.25] (6) to (3);
		\draw [style=simple, in=90, out=0, looseness=1.25] (3) to (7);
		\draw [style=simple] (14) to (15);
		\draw [style=simple] (15) to (16.center);
		\draw [style=simple] (12) to (13);
		\draw [style=simple, in=90, out=0, looseness=1.25] (13) to (14);
		\draw [style=simple] (23) to (24.center);
		\draw [style=simple, bend left=90, looseness=1.75] (20) to (21);
		\draw [style=simple] (26.center) to (27);
		\draw [style=simple] (28) to (32);
		\draw [style=simple] (32) to (33.center);
		\draw [style=simple, bend left=90, looseness=1.75] (31) to (29);
		\draw [style=simple] (34.center) to (35);
		\draw [style=simple] (36) to (40);
		\draw [style=simple] (40) to (41.center);
		\draw [style=simple] (44.center) to (45);
		\draw [style=simple] (46) to (48);
		\draw [style=simple] (48) to (49.center);
		\draw [style=simple] (10.center) to (65);
		\draw [style=simple] (65) to (14);
		\draw [style=simple, in=-180, out=90, looseness=1.25] (65) to (12);
		\draw [style=simple] (18.center) to (66);
		\draw [style=simple] (66) to (23);
		\draw [style=simple, in=-90, out=150, looseness=1.25] (66) to (20);
		\draw [style=simple, in=-90, out=30, looseness=1.25] (66) to (21);
		\draw [style=simple] (27) to (67);
		\draw [style=simple] (67) to (28);
		\draw [style=simple, in=-90, out=30, looseness=1.25] (67) to (29);
		\draw [style=simple, in=-90, out=150, looseness=1.25] (67) to (31);
		\draw [style=simple] (35) to (68);
		\draw [style=simple] (68) to (36);
		\draw [style=simple, bend left=75, looseness=1.50] (68) to (37);
		\draw [style=simple] (45) to (69);
		\draw [style=simple] (69) to (46);
		\draw [style=simple, bend left=75, looseness=1.50] (37) to (68);
		\draw [style=simple] (71.center) to (76);
		\draw [style=simple] (76) to (75);
		\draw [style=simple] (75) to (72.center);
	\end{pgfonlayer}
\end{tikzpicture}
\end{equation}

The switching function illustrates some of the peculiar properties of the use of sharing in higher-order functions. This specific instance shows how one can use self-compositions of shared functions to modify their actions, along with some intuition about what difference sharing in different bases makes.

Note that the calculuations performed in this section cannot be done using the equational rules that we have presented in this paper. Extending the equational rules to a sound system that can prove facts such as these is work for the future.
\section{Conclusion and further work}

We have presented the $\zeta$-calculus, a formal system for denoting quantum operations, with higher-order functions, where duplicated variables denote the sharing of a quantum state. We have given semantics in terms of the ZX-calculus and provided a sound equational theory. The calculus has a notion of substitution, which together with the commutation rules of ZX provides useful rewrite rules.
We have shown examples of how the abstraction mechanism of $\zeta$ can be used to represent ZX-diagrams, including phase gadgets and a linking function. We also showed instances of higher-order functions in $\zeta$ and their denotation as ZX-diagrams.

For future work, we want to expand the set of equational rules so that the behaviour of higher-order functions such as the Pauli switching functions can be catputred.

We note also that very little of the $\zeta$-calculus is depended on the fact that we are working with the bases $Z$ and $X$ for a complex vector space. The typing rules for the calculus could easily be generated to any set of bases over any vector space. 

\newpage
We have started investigating different versions of the $\zeta$-calculus for different \emph{orders of computation}, denoted $\odal_n$:

\begin{itemize}
    \item $\odal_0$ --- If we apply the $\zeta$-calculus to the one basis $\{ 0, 1 \}$ for the $\mathbb{Z}_2$-vector space $\mathbb{Z}_2^2$, we obtain the familiar $\lambda$-calculus, which we know can be applied to \emph{classical computation}.
    \item $\odal_1$ --- We can apply the $\zeta$-calculus to real vector spaces $\mathbb{R}^2$, manipulating states on an axis defined by a single spider. 
    \item $\odal_2$ --- We can apply the $\zeta$-calculus to the bases $Z$ and $X$ for $\mathbb{C}^2$ to obtain the system presented in this paper suitable for representing \emph{quantum computation}.
    \item $\odal_3$ --- We can apply the $\zeta$-calculus to bases $T$, $X$, $Y$, $Z$, ($W$) for the quaternionic vector space $\mathbb{H}^2$ to obtain a calculus suitable for calculations involving Dirac spinors, suggesting the exciting possibility of \emph{spacetime computation}. 
\end{itemize}

For each order $\zeta$-abstractions on the form $\beta^\pi x x \;:\; \underline{1} \to \underline{1}$ seem to have some connection to the generators of the different levels of the Clifford hierarchy\cite{Hiley2011ProcessFormalism}. These would be the Pauli matrices in $\odal_2$ and the $\gamma$-matrices \cite{Weinberg1995TheFields} in $\odal_3$. We hope that this will aid in revealing deeper connections between the different orders. In any case, each order extends the previous, and we hope that the correspondences between levels will help with analysing and reasoning about these different forms of computation and their commonalities.

\section*{Acknowledgements}

We would like to extend our deepest appreciation to our supervisor Robin Adams for very helpful discussions about the semantics of the system and support with editing the paper. The theory presented here makes up the foundation of the Master's thesis work on the $\zeta$-calculus both authors are currently writing.

\bibliographystyle{eptcs}
\bibliography{references}

\cleardoublepage
\appendix
\setcounter{page}{1}
\pagenumbering{Roman}			

\newpage
\section{The string diagram language} \label{ax:zx}

\begin{figure}[h]
\centering
\begin{mathpar}
\begin{tikzpicture}
	\begin{pgfonlayer}{nodelayer}
		\node [style=Tp] (0) at (0.25, 0.25) {$\alpha$};
		\node [style=Tp] (1) at (1.75, -2) {$\theta$};
		\node [style=none] (2) at (-1.5, 0.75) {};
		\node [style=none] (3) at (-1.5, -0.25) {};
		\node [style=none] (4) at (-1.5, -1.5) {};
		\node [style=none] (5) at (-1.5, -2.5) {};
		\node [style=none] (6) at (3.5, 0.75) {};
		\node [style=none] (7) at (3.5, -0.25) {};
		\node [style=none] (8) at (3.5, -1.5) {};
		\node [style=none] (9) at (3.5, -2.5) {};
		\node [style=none] (10) at (-1.25, 0.5) {$\vdots$};
		\node [style=none] (11) at (3, 0.5) {$\vdots$};
		\node [style=none] (12) at (-1, -1.75) {$\vdots$};
		\node [style=none] (13) at (3, -1.75) {$\vdots$};
		\node [style=none] (14) at (0.75, -0.5) {$\vdots$};
		\node [style=none] (15) at (5.5, -1) {$\overset{(f)}{=}$};
		\node [style=none] (16) at (7.5, 0) {};
		\node [style=none] (17) at (7.5, -2) {};
		\node [style=none] (18) at (10.5, 0) {};
		\node [style=none] (19) at (10.5, -2) {};
		\node [style=Tp] (20) at (9, -1) {$\alpha+\theta$};
		\node [style=none] (24) at (7.75, -0.75) {$\vdots$};
		\node [style=none] (25) at (10.25, -0.75) {$\vdots$};
	\end{pgfonlayer}
	\begin{pgfonlayer}{edgelayer}
		\draw [style=simple, in=165, out=0, looseness=1.25] (2.center) to (0);
		\draw [style=simple, in=-165, out=0, looseness=1.25] (3.center) to (0);
		\draw [style=simple, in=180, out=15, looseness=1.25] (0) to (6.center);
		\draw [style=simple, in=180, out=-15, looseness=1.25] (0) to (7.center);
		\draw [style=simple, in=0, out=165, looseness=1.25] (1) to (4.center);
		\draw [style=simple, in=195, out=0, looseness=1.25] (5.center) to (1);
		\draw [style=simple, in=-180, out=15, looseness=1.25] (1) to (8.center);
		\draw [style=simple, in=-180, out=-30, looseness=1.25] (1) to (9.center);
		\draw [style=simple, bend left] (1) to (0);
		\draw [style=simple, bend left] (0) to (1);
		\draw [style=simple, bend left=15] (0) to (1);
		\draw [style=simple, in=165, out=0, looseness=1.25] (16.center) to (20);
		\draw [style=simple, in=180, out=15, looseness=1.25] (20) to (18.center);
		\draw [style=simple, in=180, out=-15, looseness=1.25] (20) to (19.center);
		\draw [style=simple, in=0, out=-165, looseness=1.25] (20) to (17.center);
	\end{pgfonlayer}
\end{tikzpicture} \and
\begin{tikzpicture}
	\begin{pgfonlayer}{nodelayer}
		\node [style=none] (16) at (-1.25, 1) {};
		\node [style=none] (17) at (-1.25, -1) {};
		\node [style=none] (18) at (1.75, 1) {};
		\node [style=none] (19) at (1.75, -1) {};
		\node [style=none] (24) at (-1.75, 0.25) {$\vdots$};
		\node [style=none] (25) at (2.25, 0.25) {$\vdots$};
		\node [style=none] (30) at (-2.5, 1) {};
		\node [style=none] (31) at (-2.5, -1) {};
		\node [style=none] (32) at (3, 1) {};
		\node [style=none] (33) at (3, -1) {};
		\node [style=none] (35) at (4.5, 0) {$\overset{(h)}{=}$};
		\node [style=none] (36) at (6, 1) {};
		\node [style=none] (37) at (6, -1) {};
		\node [style=none] (38) at (9, 1) {};
		\node [style=none] (39) at (9, -1) {};
		\node [style=none] (42) at (6.25, 0.25) {$\vdots$};
		\node [style=none] (43) at (8.75, 0.25) {$\vdots$};
		\node [style=Zp] (45) at (0.25, 0) {$\alpha$};
		\node [style=H] (46) at (-1.75, 1) {};
		\node [style=H] (47) at (-1.75, -1) {};
		\node [style=H] (48) at (2.25, 1) {};
		\node [style=H] (49) at (2.25, -1) {};
		\node [style=Xp] (50) at (7.5, 0) {$\alpha$};
	\end{pgfonlayer}
	\begin{pgfonlayer}{edgelayer}
		\draw [style=simple] (30.center) to (46);
		\draw [style=simple] (46) to (16.center);
		\draw [style=simple] (31.center) to (47);
		\draw [style=simple] (47) to (17.center);
		\draw [style=simple] (18.center) to (48);
		\draw [style=simple] (48) to (32.center);
		\draw [style=simple] (19.center) to (49);
		\draw [style=simple] (49) to (33.center);
		\draw [style=simple, in=165, out=0, looseness=1.25] (16.center) to (45);
		\draw [style=simple, in=180, out=15, looseness=1.25] (45) to (18.center);
		\draw [style=simple, in=0, out=-165, looseness=1.25] (45) to (17.center);
		\draw [style=simple, in=180, out=-15, looseness=1.25] (45) to (19.center);
		\draw [style=simple, in=165, out=0, looseness=1.25] (36.center) to (50);
		\draw [style=simple, in=-180, out=15, looseness=1.25] (50) to (38.center);
		\draw [style=simple, in=180, out=-15, looseness=1.25] (50) to (39.center);
		\draw [style=simple, in=0, out=-165, looseness=1.25] (50) to (37.center);
	\end{pgfonlayer}
\end{tikzpicture} \and
\begin{tikzpicture}
	\begin{pgfonlayer}{nodelayer}
		\node [style=none] (1) at (1, 0) {};
		\node [style=none] (2) at (-1, 0) {};
		\node [style=T] (4) at (0, 0) {};
		\node [style=none] (5) at (2, 0.25) {$\overset{(id)}{=}$};
		\node [style=none] (6) at (5, 0) {};
		\node [style=none] (7) at (3, 0) {};
	\end{pgfonlayer}
	\begin{pgfonlayer}{edgelayer}
		\draw [style=simple] (2.center) to (4);
		\draw [style=simple] (4) to (1.center);
		\draw [style=simple] (7.center) to (6.center);
	\end{pgfonlayer}
\end{tikzpicture} \and
\begin{tikzpicture}
	\begin{pgfonlayer}{nodelayer}
		\node [style=none] (4) at (1, 0) {};
		\node [style=none] (5) at (-2, 0) {};
		\node [style=none] (14) at (2, 0.25) {$\overset{(hh)}{=}$};
		\node [style=none] (16) at (3.25, 0) {};
		\node [style=none] (17) at (5, 0) {};
		\node [style=H] (18) at (-1, 0) {};
		\node [style=H] (19) at (0, 0) {};
	\end{pgfonlayer}
	\begin{pgfonlayer}{edgelayer}
		\draw [style=simple] (16.center) to (17.center);
		\draw (5.center) to (18);
		\draw (18) to (19);
		\draw (19) to (4.center);
	\end{pgfonlayer}
\end{tikzpicture} \and
\begin{tikzpicture}
	\begin{pgfonlayer}{nodelayer}
		\node [style=Zp] (0) at (0, 0) {$\alpha$};
		\node [style=Xp] (1) at (-1.5, 0) {$a \pi$};
		\node [style=none] (2) at (1.5, 1) {};
		\node [style=none] (3) at (1.5, 0) {};
		\node [style=none] (4) at (1.5, -1) {};
		\node [style=none] (7) at (1.25, -0.25) {$\vdots$};
		\node [style=none] (8) at (2.75, 0.25) {$\overset{(\pi)}{=}$};
		\node [style=Xp] (9) at (7.5, 1.25) {$a \pi$};
		\node [style=Xp] (10) at (7.5, 0) {$a \pi$};
		\node [style=Xp] (11) at (7.5, -1.25) {$a \pi$};
		\node [style=none] (15) at (8.75, 1.25) {};
		\node [style=none] (16) at (8.75, 0) {};
		\node [style=none] (17) at (8.75, -1.25) {};
		\node [style=none] (19) at (-3, 0) {};
		\node [style=Zp] (20) at (5.5, 0) {$-\alpha$};
		\node [style=none] (22) at (7, 1.25) {};
		\node [style=none] (23) at (7, 0) {};
		\node [style=none] (24) at (7, -1.25) {};
		\node [style=none] (27) at (8.25, -0.5) {$\vdots$};
		\node [style=none] (28) at (4, 0) {};
	\end{pgfonlayer}
	\begin{pgfonlayer}{edgelayer}
		\draw [style=simple] (1) to (0);
		\draw [style=simple, in=-180, out=15, looseness=1.25] (0) to (2.center);
		\draw [style=simple] (0) to (3.center);
		\draw [style=simple, in=180, out=-15, looseness=1.25] (0) to (4.center);
		\draw [style=simple] (9) to (15.center);
		\draw [style=simple] (10) to (16.center);
		\draw [style=simple] (11) to (17.center);
		\draw [style=simple] (19.center) to (1);
		\draw [style=simple, in=-180, out=15, looseness=1.25] (20) to (22.center);
		\draw [style=simple] (20) to (23.center);
		\draw [style=simple, in=180, out=-15, looseness=1.25] (20) to (24.center);
		\draw [style=simple] (28.center) to (20);
		\draw [style=simple] (22.center) to (9);
		\draw [style=simple] (23.center) to (10);
		\draw [style=simple] (24.center) to (11);
	\end{pgfonlayer}
\end{tikzpicture} \and
\begin{tikzpicture}
	\begin{pgfonlayer}{nodelayer}
		\node [style=Zp] (0) at (0, 0) {$\alpha$};
		\node [style=Xp] (1) at (-1.5, 0) {$a \pi$};
		\node [style=none] (2) at (1.5, 1) {};
		\node [style=none] (3) at (1.5, 0) {};
		\node [style=none] (4) at (1.5, -1) {};
		\node [style=none] (7) at (1.25, -0.25) {$\vdots$};
		\node [style=none] (8) at (2.75, 0) {$\overset{(c)}{=}$};
		\node [style=Xp] (9) at (5, 1.25) {$a \pi$};
		\node [style=Xp] (10) at (5, 0) {$a \pi$};
		\node [style=Xp] (11) at (5, -1.25) {$a \pi$};
		\node [style=none] (15) at (6.75, 1.25) {};
		\node [style=none] (16) at (6.75, 0) {};
		\node [style=none] (17) at (6.75, -1.25) {};
		\node [style=none] (18) at (6.5, -0.5) {$\vdots$};
	\end{pgfonlayer}
	\begin{pgfonlayer}{edgelayer}
		\draw [style=simple] (1) to (0);
		\draw [style=simple, in=-180, out=15, looseness=1.25] (0) to (2.center);
		\draw [style=simple] (0) to (3.center);
		\draw [style=simple, in=180, out=-15, looseness=1.25] (0) to (4.center);
		\draw [style=simple] (9) to (15.center);
		\draw [style=simple] (10) to (16.center);
		\draw [style=simple] (11) to (17.center);
	\end{pgfonlayer}
\end{tikzpicture} \and
\begin{tikzpicture}
	\begin{pgfonlayer}{nodelayer}
		\node [style=X] (0) at (0, 0) {};
		\node [style=Z] (1) at (1, 0) {};
		\node [style=none] (2) at (-1, 0.75) {};
		\node [style=none] (3) at (-1, -0.75) {};
		\node [style=none] (4) at (2, 0.75) {};
		\node [style=none] (5) at (2, -0.75) {};
		\node [style=none] (6) at (-1, 0.25) {$\vdots$};
		\node [style=none] (7) at (2, 0.25) {$\vdots$};
		\node [style=none] (8) at (-1.5, -0.75) {};
		\node [style=none] (9) at (-1.5, 0.75) {};
		\node [style=none] (10) at (2.5, 0.75) {};
		\node [style=none] (11) at (2.5, -0.75) {};
		\node [style=none] (12) at (-2, 0) {$n$};
		\node [style=none] (13) at (3, 0) {$m$};
		\node [style=none] (16) at (4, 0.25) {$\overset{(b)}{=}$};
		\node [style=none] (19) at (6, 0.75) {};
		\node [style=none] (20) at (6, -0.75) {};
		\node [style=none] (21) at (9, 0.75) {};
		\node [style=none] (22) at (9, -0.75) {};
		\node [style=none] (23) at (6, 0.25) {$\vdots$};
		\node [style=none] (24) at (9, 0.25) {$\vdots$};
		\node [style=none] (25) at (5.5, -0.75) {};
		\node [style=none] (26) at (5.5, 0.75) {};
		\node [style=none] (27) at (9.5, 0.75) {};
		\node [style=none] (28) at (9.5, -0.75) {};
		\node [style=none] (29) at (5, 0) {$n$};
		\node [style=none] (30) at (10, 0) {$m$};
		\node [style=Z] (31) at (6.75, 0.75) {};
		\node [style=Z] (32) at (6.75, -0.75) {};
		\node [style=X] (33) at (8.25, 0.75) {};
		\node [style=X] (34) at (8.25, -0.75) {};
	\end{pgfonlayer}
	\begin{pgfonlayer}{edgelayer}
		\draw [style=simple, in=165, out=0, looseness=1.25] (2.center) to (0);
		\draw [style=simple, in=0, out=-165, looseness=1.25] (0) to (3.center);
		\draw [style=simple] (0) to (1);
		\draw [style=simple, in=15, out=180, looseness=1.25] (4.center) to (1);
		\draw [style=simple, in=180, out=-15, looseness=1.25] (1) to (5.center);
		\draw [style=brace edge] (8.center) to (9.center);
		\draw [style=brace edge] (10.center) to (11.center);
		\draw [style=brace edge] (25.center) to (26.center);
		\draw [style=brace edge] (27.center) to (28.center);
		\draw [style=simple] (19.center) to (31);
		\draw [style=simple] (20.center) to (32);
		\draw [style=simple] (33) to (21.center);
		\draw [style=simple] (34) to (22.center);
		\draw [style=simple] (31) to (33);
		\draw [style=simple] (31) to (34);
		\draw [style=simple] (32) to (34);
		\draw [style=simple] (32) to (33);
	\end{pgfonlayer}
\end{tikzpicture}
\end{mathpar}
\caption{The equational rules of the string diagram language.}
\label{fig:string_diagram_rules}
\end{figure}
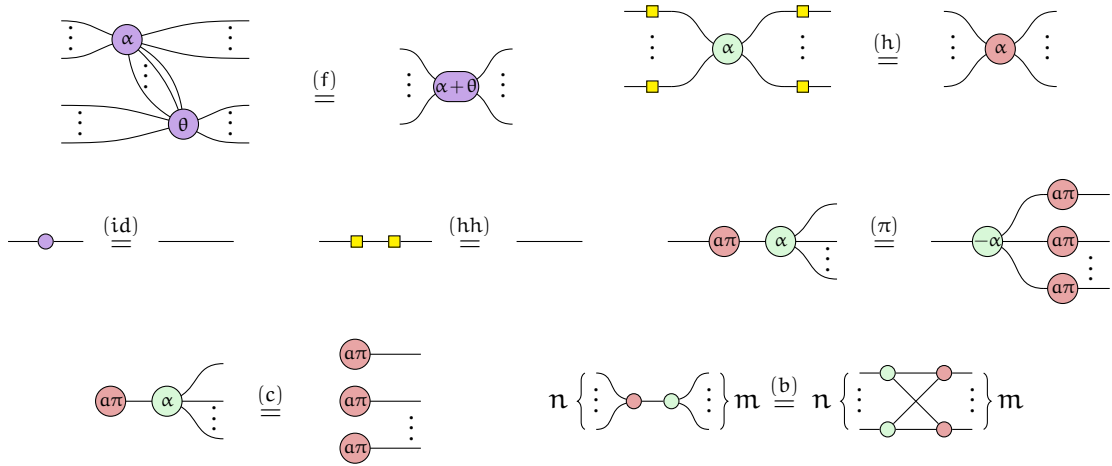

The string diagram language is, essentially, the ZX-calculus\cite{vandeWetering2020ZX-calculusScientist}, with the only modification being that we denote a general spider by a purple node. The semantics and equational theory of the string diagrams is described in figures \ref{fig:denotational_semantics} and \ref{fig:string_diagram_rules}. When naming the basis of a spider in a diagram, we hold the following conventions on their names and colours.
\begin{itemize}
    \item $\beta$ is a general basis, and is purple.
    \item $\zeta$ and $\xi$ are the $Z$ and $X$ bases, with their usual colours.
\end{itemize}
\begin{figure}[h] 
\begin{mathpar}
\left\llbracket\; \begin{tikzpicture}
	\begin{pgfonlayer}{nodelayer}
		\node [style=none] (0) at (-0.75, 0) {};
		\node [style=none] (1) at (0.75, 0) {};
	\end{pgfonlayer}
	\begin{pgfonlayer}{edgelayer}
		\draw [style=simple] (0.center) to (1.center);
	\end{pgfonlayer}
\end{tikzpicture} \;\right\rrbracket
:=
\ketbra{0} + \ketbra{1}
\and
\left\llbracket \begin{tikzpicture}
	\begin{pgfonlayer}{nodelayer}
		\node [style=none] (0) at (0, 0.5) {};
		\node [style=none] (1) at (0, -0.5) {};
		\node [style=none] (2) at (0.5, 0.5) {};
		\node [style=none] (3) at (0.5, -0.5) {};
	\end{pgfonlayer}
	\begin{pgfonlayer}{edgelayer}
		\draw [style=simple, bend right=90, looseness=1.75] (0.center) to (1.center);
		\draw [style=simple] (0.center) to (2.center);
		\draw [style=simple] (1.center) to (3.center);
	\end{pgfonlayer}
\end{tikzpicture} \;\right\rrbracket
:=
\ket{00} + \ket{11}
\and
\left\llbracket\; \begin{tikzpicture}
	\begin{pgfonlayer}{nodelayer}
		\node [style=none] (0) at (0, 0.5) {};
		\node [style=none] (1) at (0, -0.5) {};
		\node [style=none] (2) at (-0.5, 0.5) {};
		\node [style=none] (3) at (-0.5, -0.5) {};
	\end{pgfonlayer}
	\begin{pgfonlayer}{edgelayer}
		\draw [style=simple, bend left=90, looseness=1.75] (0.center) to (1.center);
		\draw [style=simple] (0.center) to (2.center);
		\draw [style=simple] (1.center) to (3.center);
	\end{pgfonlayer}
\end{tikzpicture} \right\rrbracket
:=
\bra{00} + \bra{11}
\and
\left\llbracket\; \begin{tikzpicture}
	\begin{pgfonlayer}{nodelayer}
		\node [style=none] (1) at (-1, 0.5) {};
		\node [style=none] (2) at (-1, -0.5) {};
		\node [style=none] (3) at (1, 0.5) {};
		\node [style=none] (4) at (1, -0.5) {};
	\end{pgfonlayer}
	\begin{pgfonlayer}{edgelayer}
		\draw [in=180, out=0, looseness=1.50] (1.center) to (4.center);
		\draw [in=180, out=0, looseness=1.50] (2.center) to (3.center);
	\end{pgfonlayer}
\end{tikzpicture} \;\right\rrbracket
:=
\ketbra{00}{00} + \ketbra{01}{10} + \ketbra{10}{01} + \ketbra{11}{11}
\and
\left\llbracket\; \begin{tikzpicture}
	\begin{pgfonlayer}{nodelayer}
		\node [style=none] (1) at (-1.5, 0) {};
		\node [style=none] (2) at (1.5, 0) {};
		\node [style=H] (3) at (0, 0) {};
	\end{pgfonlayer}
	\begin{pgfonlayer}{edgelayer}
		\draw [style=simple] (1.center) to (2.center);
	\end{pgfonlayer}
\end{tikzpicture} \;\right\rrbracket
:=
\ketbra{+}{0} + \ketbra{-}{1}
\and
\left\llbracket \begin{tikzpicture}
	\begin{pgfonlayer}{nodelayer}
		\node [style=none] (0) at (1.5, 1) {};
		\node [style=none] (1) at (1.5, -1) {};
		\node [style=none] (2) at (5.5, 1) {};
		\node [style=none] (3) at (5.5, -1) {};
		\node [style=none] (4) at (1, 1) {};
		\node [style=none] (5) at (1, -1) {};
		\node [style=none] (6) at (0.5, 0) {$m$};
		\node [style=none] (7) at (6, 1) {};
		\node [style=none] (8) at (6, -1) {};
		\node [style=none] (9) at (6.5, 0) {$n$};
		\node [style=none] (11) at (5.25, 0.25) {$\vdots$};
		\node [style=none] (12) at (1.75, 0.25) {$\vdots$};
		\node [style=Tp] (13) at (3.5, 0) {$\theta$};
	\end{pgfonlayer}
	\begin{pgfonlayer}{edgelayer}
		\draw [style=brace edge] (7.center) to (8.center);
		\draw [style=brace edge] (5.center) to (4.center);
		\draw [style=simple, in=165, out=0, looseness=1.25] (0.center) to (13);
		\draw [style=simple, in=-180, out=15, looseness=1.25] (13) to (2.center);
		\draw [style=simple, in=-15, out=-180, looseness=1.25] (3.center) to (13);
		\draw [style=simple, in=0, out=-165, looseness=1.25] (13) to (1.center);
	\end{pgfonlayer}
\end{tikzpicture} \right\rrbracket
:=
\ket{\beta_0}^{\otimes n} \bra{\beta_0}^{\otimes m} + e^{i \theta} \ket{\beta_1}^{\otimes n} \bra{\beta_1}^{\otimes m}
\end{mathpar}

\caption{Denotational semantics of the string diagram language.}
\label{fig:denotational_semantics}
\end{figure}
\section{Soundness of the equational theory}
\label{appendix:soundness}

\begin{mathpar}
\resizebox{\textwidth}{!}{\input{tikz/soundness.tikz}}
\end{mathpar}

\newpage
\section{Derivation of higher-order sharing} \label{appendix:sharing_sharing}

\vfill

\begin{mathpar}
\inferrule*[right=a]
    { \inferrule*[right=b]
        { \inferrule*[right=t]
            { \inferrule*[right=v]
                { f :_\xi\! A \to A \in \{f :_\xi\! A \to A\}}
                {f :_\xi\! A \to A \otimes A \vdash f : A \to A \otimes A}
            \\ \inferrule*[right=v]
                {  f :_\xi\! A \to A \in \{f :_\xi\! A \to A\}}
                {f :_\xi\! A \to A \otimes A \vdash f : A \to A \otimes A}
            }
            { f :_\xi\! A \to A \otimes A \vdash \langle f, f \rangle : (A \to A \otimes A) \otimes (A \to A \otimes A) }
        }
        { \vdash \xi f \langle f, f \rangle : (A \to A \otimes A) \to (A \to A \otimes A) \otimes (A \to A \otimes A) }
    \\ \inferrule*[right=b]
        { \inferrule*[right=t]
            { \inferrule*[right=v]
                { x :_\zeta\! A \in \{ x :_\zeta\! A \} }
                { x :_\zeta\! A \vdash x : A }
            \\ \inferrule*[right=v]
                {  x :_\zeta\! A \in \{ x :_\zeta\! A \} }
                { x :_\zeta\! A \vdash x : A }
            }
            { x :_\zeta\! A \vdash \langle x, x \rangle : A \otimes A }
        }
        { \vdash \zeta x \langle x, x \rangle : A \to A \otimes A }
    }
    { \vdash (\xi f \langle f, f \rangle) (\zeta x \langle x, x \rangle) : (A \to A \otimes A) \otimes (A \to A \otimes A) }
\end{mathpar}

\vfill

\begin{center}
\resizebox{\textwidth}{!}{\begin{tikzpicture}
	\begin{pgfonlayer}{nodelayer}
		\node [style=none] (1) at (8, 4.5) {};
		\node [style=none] (3) at (8, 2.5) {};
		\node [style=none] (5) at (6.25, 3.25) {};
		\node [style=none] (6) at (4.75, 3.25) {};
		\node [style=none] (7) at (4.75, 2) {};
		\node [style=none] (8) at (6.25, 2) {};
		\node [style=none] (9) at (3, 5) {$A^*$};
		\node [style=none] (10) at (3, 3) {$A^*$};
		\node [style=none] (12) at (5.5, 7.25) {};
		\node [style=none] (13) at (8, 6) {};
		\node [style=none] (14) at (8, 7) {};
		\node [style=none] (16) at (3, 7) {$A$};
		\node [style=none] (17) at (10.25, 8) {};
		\node [style=none] (18) at (10.25, 1.5) {};
		\node [style=none] (19) at (1.5, 1.5) {};
		\node [style=none] (20) at (1.5, 8) {};
		\node [style=none] (21) at (-4.75, 0) {};
		\node [style=none] (22) at (12.5, 0) {};
		\node [style=none] (23) at (-4.75, -1) {};
		\node [style=none] (24) at (12.5, -1) {};
		\node [style=none] (25) at (-4, 0.5) {};
		\node [style=none] (26) at (-2.5, 0.5) {};
		\node [style=none] (27) at (-2.5, -2.5) {};
		\node [style=none] (28) at (-4, -2.5) {};
		\node [style=none] (32) at (-3.25, 5.75) {};
		\node [style=none] (33) at (-4.75, 6) {};
		\node [style=none] (34) at (-4.75, 7) {};
		\node [style=none] (37) at (-3.25, 3.75) {};
		\node [style=none] (38) at (-4.75, 4) {};
		\node [style=none] (39) at (-4.75, 5) {};
		\node [style=none] (42) at (-3.25, 1.75) {};
		\node [style=none] (43) at (-4.75, 2) {};
		\node [style=none] (44) at (-4.75, 3) {};
		\node [style=none] (46) at (-4.75, -2) {};
		\node [style=none] (47) at (12.5, -2) {};
		\node [style=none] (48) at (-4.75, -3.75) {};
		\node [style=none] (49) at (12.5, -3.75) {};
		\node [style=none] (50) at (-4.75, -4.75) {};
		\node [style=none] (51) at (12.5, -4.75) {};
		\node [style=none] (52) at (-4, -3.25) {};
		\node [style=none] (53) at (-2.5, -3.25) {};
		\node [style=none] (54) at (-2.5, -6.25) {};
		\node [style=none] (55) at (-4, -6.25) {};
		\node [style=none] (56) at (-4.75, -5.75) {};
		\node [style=none] (57) at (12.5, -5.75) {};
		\node [style=none] (58) at (13, 0) {$A^*$};
		\node [style=none] (59) at (13, -1) {$A$};
		\node [style=none] (60) at (13, -2) {$A$};
		\node [style=none] (61) at (13, -3.75) {$A^*$};
		\node [style=none] (62) at (13, -4.75) {$A$};
		\node [style=none] (63) at (13, -5.75) {$A$};
		\node [style=none] (64) at (-12.25, 8) {};
		\node [style=none] (65) at (0.75, 8) {};
		\node [style=none] (66) at (0.75, -7.25) {};
		\node [style=none] (67) at (-12.25, -7.25) {};
		\node [style=none] (120) at (6.25, 5) {};
		\node [style=none] (121) at (4.75, 5) {};
		\node [style=none] (122) at (4.75, 3.75) {};
		\node [style=none] (123) at (6.25, 3.75) {};
		\node [style=none] (124) at (6.75, 5.5) {};
		\node [style=none] (125) at (4.25, 5.5) {};
		\node [style=none] (126) at (4.25, 1.75) {};
		\node [style=none] (127) at (6.75, 1.75) {};
		\node [style=none] (128) at (-2, 1) {};
		\node [style=none] (129) at (-4.5, 1) {};
		\node [style=none] (130) at (-4.5, -6.75) {};
		\node [style=none] (131) at (-2, -6.75) {};
		\node [style=none] (132) at (-13, 8.75) {};
		\node [style=none] (133) at (11, 8.75) {};
		\node [style=none] (134) at (-13, -8) {};
		\node [style=none] (135) at (11, -8) {};
		\node [style=none] (137) at (-6, 6) {};
		\node [style=none] (139) at (-6, 7) {};
		\node [style=none] (140) at (10, 7.75) {{\tiny \textbf{B}$^\dagger$}};
		\node [style=none] (141) at (-7, 5) {};
		\node [style=none] (142) at (-7, 4) {};
		\node [style=none] (143) at (-8, 3) {};
		\node [style=none] (144) at (-8, 2) {};
		\node [style=none] (145) at (-6, 0) {};
		\node [style=none] (146) at (-6, -3.75) {};
		\node [style=none] (147) at (-7, -1) {};
		\node [style=none] (148) at (-7, -4.75) {};
		\node [style=none] (149) at (-8, -2) {};
		\node [style=none] (150) at (-8, -5.75) {};
		\node [style=Z] (151) at (5.5, 6.5) {};
		\node [style=X] (152) at (-3.25, 6.5) {};
		\node [style=X] (153) at (-3.25, 4.5) {};
		\node [style=X] (154) at (-3.25, 2.5) {};
		\node [style=none] (155) at (-2.75, 0.25) {{\tiny \textbf{V}}};
		\node [style=none] (156) at (-2.75, -3.5) {{\tiny \textbf{V}}};
		\node [style=none] (157) at (-2.25, 0.75) {{\tiny \textbf{T}}};
		\node [style=none] (158) at (6, 3) {{\tiny \textbf{V}}};
		\node [style=none] (159) at (6, 4.75) {{\tiny \textbf{V}}};
		\node [style=none] (160) at (6.5, 5.25) {{\tiny \textbf{T}}};
		\node [style=none] (161) at (0.5, 7.75) {{\tiny \textbf{B}}};
		\node [style=none] (162) at (10.75, 8.5) {{\tiny \textbf{A}}};
	\end{pgfonlayer}
	\begin{pgfonlayer}{edgelayer}
		\draw [style=simple] (6.center)
			 to (7.center)
			 to (8.center)
			 to (5.center)
			 to cycle;
		\draw [style=simple, bend left=90, looseness=1.50] (13.center) to (1.center);
		\draw [style=simple, bend left=90, looseness=1.50] (14.center) to (3.center);
		\draw [style=simple] (20.center)
			 to (19.center)
			 to (18.center)
			 to (17.center)
			 to cycle;
		\draw [style=simple] (21.center) to (22.center);
		\draw [style=simple] (23.center) to (24.center);
		\draw [style=simple] (26.center)
			 to (27.center)
			 to (28.center)
			 to (25.center)
			 to cycle;
		\draw [style=simple] (46.center) to (47.center);
		\draw [style=simple] (48.center) to (49.center);
		\draw [style=simple] (50.center) to (51.center);
		\draw [style=simple] (55.center)
			 to (52.center)
			 to (53.center)
			 to (54.center)
			 to cycle;
		\draw [style=simple] (56.center) to (57.center);
		\draw [style=simple] (67.center)
			 to (64.center)
			 to (65.center)
			 to (66.center)
			 to cycle;
		\draw [style=simple] (121.center)
			 to (122.center)
			 to (123.center)
			 to (120.center)
			 to cycle;
		\draw [style=simple] (125.center)
			 to (126.center)
			 to (127.center)
			 to (124.center)
			 to cycle;
		\draw [style=simple] (131.center)
			 to (128.center)
			 to (129.center)
			 to (130.center)
			 to cycle;
		\draw (132.center)
			 to (133.center)
			 to (135.center)
			 to (134.center)
			 to cycle;
		\draw [bend right=90, looseness=1.50] (139.center) to (145.center);
		\draw [bend left=270, looseness=1.50] (137.center) to (146.center);
		\draw [bend right=90, looseness=1.50] (141.center) to (147.center);
		\draw [bend right=90, looseness=1.50] (142.center) to (148.center);
		\draw [bend right=90, looseness=1.75] (143.center) to (149.center);
		\draw [bend left=270, looseness=1.50] (144.center) to (150.center);
		\draw (139.center) to (34.center);
		\draw (137.center) to (33.center);
		\draw (141.center) to (39.center);
		\draw (142.center) to (38.center);
		\draw (143.center) to (44.center);
		\draw (144.center) to (43.center);
		\draw (145.center) to (21.center);
		\draw (147.center) to (23.center);
		\draw (149.center) to (46.center);
		\draw (146.center) to (48.center);
		\draw (148.center) to (50.center);
		\draw (150.center) to (56.center);
		\draw [style=simple, in=165, out=0, looseness=1.25] (34.center) to (152);
		\draw [style=simple, in=0, out=-165, looseness=1.25] (152) to (33.center);
		\draw [style=simple, in=165, out=0, looseness=1.25] (39.center) to (153);
		\draw [style=simple, in=0, out=-165, looseness=1.25] (153) to (38.center);
		\draw [style=simple, in=165, out=0, looseness=1.25] (44.center) to (154);
		\draw [style=simple, in=0, out=-165, looseness=1.25] (154) to (43.center);
		\draw [style=simple] (152) to (151);
		\draw [style=simple] (153) to (1.center);
		\draw [style=simple] (154) to (3.center);
		\draw [style=simple, in=180, out=15, looseness=1.25] (151) to (14.center);
		\draw [style=simple, in=-15, out=180, looseness=1.25] (13.center) to (151);
	\end{pgfonlayer}
\end{tikzpicture}}
\end{center}

\vfill

\[
\left\llbracket\; \begin{tikzpicture}
	\begin{pgfonlayer}{nodelayer}
		\node [style=none] (3) at (0, -0.75) {};
		\node [style=none] (7) at (-1.5, -0.75) {};
		\node [style=none] (8) at (-3, -0.5) {};
		\node [style=none] (9) at (-3, 0.5) {};
		\node [style=none] (11) at (1.5, 1.75) {};
		\node [style=none] (12) at (3, -0.5) {};
		\node [style=none] (13) at (3, 1.5) {};
		\node [style=none] (15) at (1.5, -1.75) {};
		\node [style=none] (16) at (3, -1.5) {};
		\node [style=none] (17) at (3, 0.5) {};
		\node [style=Z] (26) at (-1.5, 0) {};
		\node [style=Z] (27) at (1.5, 1) {};
		\node [style=Z] (28) at (1.5, -1) {};
		\node [style=X] (29) at (0, 0) {};
		\node [style=none] (32) at (0.25, 0) {};
		\node [style=Z] (33) at (0, 0) {};
		\node [style=X] (34) at (1.5, 1) {};
		\node [style=X] (35) at (1.5, -1) {};
		\node [style=X] (36) at (-1.5, 0) {};
	\end{pgfonlayer}
	\begin{pgfonlayer}{edgelayer}
		\draw [style=simple, in=165, out=0] (9.center) to (26);
		\draw [style=simple, in=0, out=-165] (26) to (8.center);
		\draw [style=simple] (26) to (29);
		\draw [style=simple, in=180, out=15] (29) to (27);
		\draw [style=simple, in=180, out=-15] (29) to (28);
		\draw [style=simple, in=180, out=15] (27) to (13.center);
		\draw [style=simple, in=180, out=-30] (27) to (12.center);
		\draw [style=simple, in=180, out=15] (28) to (17.center);
		\draw [style=simple, in=-180, out=-15] (28) to (16.center);
	\end{pgfonlayer}
\end{tikzpicture} \;\right\rrbracket
\quad=\quad
\begin{matrix}
\left( \mathbb{1} \otimes \sigma \otimes \mathbb{1} \right) \\
\left( \ketbra{++}{+}+\ketbra{--}{-} \right)^{\otimes 2} \\
\left( \ketbra{00}{0}+\ketbra{11}{1} \right) \\
\left( \ketbra{+}{++}+\ketbra{-}{--} \right)
\end{matrix}
\]

\end{document}